\documentclass[aps,prd,showpacs,nofootinbib,twocolumn]{revtex4}
\usepackage{latexsym}
\usepackage{amsmath,amsfonts}
\usepackage{amsbsy}
\usepackage{mathrsfs}
\usepackage{color}
\usepackage{float}
\usepackage{dsfont}
\usepackage{psfrag}
\usepackage{enumerate}
\usepackage{soul}
\usepackage{gensymb}

\usepackage{amsmath,amssymb,calc,amsfonts}
\usepackage{latexsym}
\usepackage{hyperref}

\usepackage{graphicx,calc,epsfig}

\begin{document}

\title{
Gravitational lensing in dispersive media and deflection angle of charged massive particles
in terms of curvature scalars and energy-momentum tensor
}
 
\date{\today}

 \author{Gabriel Crisnejo$^1$, Emanuel Gallo${^{1,2}}$ and Jos\'e R. Villanueva$^3$} 

 \affiliation{ 
 $^1$FaMAF, UNC; Ciudad Universitaria, (5000) C\'ordoba, Argentina. \\ 
 $^2$Instituto de F\'isica Enrique Gaviola (IFEG), CONICET, \\
 Ciudad Universitaria, (5000) C\'ordoba, Argentina.\\
 $^3$ Instituto de F\'isica y Astronom\'ia, Universidad de Valpara\'iso, Gran Breta\~na 1111, Valpara\'iso, Chile.}

\begin{abstract}
In this work we extend the approach used in [Emanuel Gallo and Osvaldo M. Moreschi, Phys. Rev. D 83,
12 083007 (2011)] to the study of weak gravitational lensing in a plasma medium. First, we present expressions for the deflection angle and optical scalars in terms of the components of the energy-momentum tensor for spherically symmetric lenses surrounded by a cold nonmagnetized plasma. Second, we show that the same expressions can be deduced using the Gauss-Bonnet theorem. Finally, we establish a correspondence between the spatial orbits of photons in a nonhomogeneous plasma and the nongeodesic curves followed by test massive particles whose dynamics also depend on an external central field. As an application, we use the Gauss-Bonnet theorem to compute the deflection angle of the nongeodesic trajectories followed by relativistic test massive charged particles in a Reissner-Nordstr\"om spacetime.
\end{abstract}

\maketitle

{\sf \footnotesize \scriptsize
}

\vspace{5mm}

\section{Introduction}

Gravitational lensing is an essential tool to study the content of matter and energy in the Universe. In many observations, the influence of optical media along the path of the light rays propagation can be neglected. However, environmental effects cannot be safely neglected in the radio-frequency range. At all physical scales we are physically motivated to consider that galaxies, clusters of galaxies, black holes or other compact objects are surrounded by a dispersive plasma medium. In such an optical medium, the propagation of light rays becomes frequency-dependent.

The effect of plasma on light propagation has been studied since 1960. The influence of the solar corona on the time delay was first investigated by Muhleman and Johnston in 1966 \cite{Solar-Radio} and four years later Muhleman, Ekers and Fomalont calculated the light deflection in the presence of a plasma in the weak-field approximation \cite{Muhleman-1970}. In 1980 Breuer and Ehlers performed a rigorous derivation of a Hamiltonian for light rays including a magnetized plasma and curved background \cite{Breuer-1980, Breuer-1981a, Breuer-1981b}. The light deflection in a plasma was calculated for the first time in the Schwarzschild spacetime (and in the equatorial plane of the Kerr spacetime) without the weak-field approximation by Perlick in 2000 \cite{Perlick-book}. Since then the study of the influence of plasma media become an active research area \cite{BisnovatyiKogan:2008yg,
BisnovatyiKogan:2010ar,
Tsupko:2013cqa,Tsupko:2014sca,
Tsupko:2014lta,Perlick:2015vta,
Bisnovatyi-Kogan:2015dxa,
Perlick:2017fio,
Bisnovatyi-Kogan:2017kii, 
2013ApSS.346..513M,
2016ApSS.361..226A, 
2017IJMPD..2650051A,
2017IJMPD..2641011A,
2017PhRvD..96h4017A,
2018arXiv180203293T,
Rogers:2015dla,
Rogers:2016xcc,
Rogers:2017ofq,
2018MNRAS.475..867E, 
Crisnejo-gauss-bonnet-1,Er:2013efa,Yan:2019etp}.

One of the crucial quantities in the study of gravitational lensing is the deflection angle. In general, expressions for the deflection angle are written in terms of derivatives of the metric components. However, in \cite{Gallo-lens-2011}, Gallo and Moreschi introduced an expression for the deflection angle in the weak lensing regime which is written in terms of curvature scalars. More precisely, in the case of a static, axially symmetric lens configuration,
it was shown that the deflection angle $\alpha$ in a weak field regime can be written  at linear order in terms of the impact parameter $b$, and the projected Ricci and Weyl scalars $\hat\Phi_{00}$ and $\hat\Psi_0=-\hat\psi_0 e^{2i\vartheta}$
(we refer to \cite{Gallo-lens-2011} for more details) in a very
compact form which we reproduce here \begin{equation}\label{eq:alphaold}
    \alpha(b)=b(\hat\Phi_{00}(b)+\hat\psi_0(b)).
\end{equation}
The advantage of \eqref{eq:alphaold} is that it is written in terms of geometrical  quantities with a clear physical meaning  instead of the usual coordinate dependent expressions in terms of the metric components. 
Recently, it was generalized to the cosmological context by Boero and Moreschi \cite{Boero:2016nrd} and by us to take into account second order corrections in perturbations of a flat metric \cite{Crisnejo-lens-2018}. In addition, this approach was used to study the dark matter phenomena and alternatives to the Schwarzschild metric \cite{Gallo-peculiar,Bozza:2015haa}.

On the other hand, Gibbons and Werner have also established an alternative geometrical (and topological) way to study gravitational lensing using the Gauss-Bonnet theorem and an associated two-dimensional optical metric \cite{Gibbons-gauss-bonnet}. This method allows us to calculate the deflection angle in terms of other geometrical quantities: the Gaussian curvature of an appropriate domain $D$, the geodesic curvatures of the curves which conform the boundary $\partial D$ of $D$, and its Euler number which is a  topological invariant. Since the seminal work of Gibbons and Werner, the Gauss-Bonnet theorem has been applied in  numerous studies  \cite{Jusufi:2015laa,Jusufi:2016wiz,Jusufi:2016sym,Jusufi:2017gyu,Ovgun:2018ran,Ovgun:2018fnk,Jusufi:2018jof,Jusufi:2017uhh,Jusufi:2017mav,Jusufi:2017vta,Jusufi:2017drg,Jusufi:2017xnr,Jusufi:2018waj,Ishihara:2016vdc,Ishihara:2016sfv,Ovgun:2018xys,Ovgun:2018ran,Jusufi:2018kry,Jusufi:2017hed,Jusufi:2017lsl,Ovgun:2018tua,Javed:2019qyg, deLeon:2019qnp,Haroon:2018ryd,Jusufi:2018gnz}. In particular this method was successfully extended by us to the plasma case \cite{Crisnejo-gauss-bonnet-1,Crisnejo:2018ppm}. It is worthwhile to mention that the concept of the optical metric and the related Fermat principle for light rays in general relativity were introduced first by Weyl in 1917 \cite{Weyl-1917}.

Therefore, for the pure gravity case there are at least two alternative methods to compute the deflection angle and related optical quantities in terms of coordinate-free quantities: on the one hand the null-tetrad approach\cite{Gallo-lens-2011,Crisnejo-lens-2018}, and on the other the Gibbons-Werner method \cite{Gibbons-gauss-bonnet}. Note that even when these two methods are  geometrical, the associated geometrical quantities refer to two different manifolds (a four-dimensional Lorentzian metric in the case of \cite{Gallo-lens-2011}) and a two-dimensional Riemannian metric in the case of \cite{Gibbons-gauss-bonnet}. As we mentioned above, we have also recently extended the use of the Gibbons-Werner method to more general situations where dispersive media are present. A natural question arises: can the null tetrad method developed in \cite{Gallo-lens-2011} also be extended to apply to dispersive media? It is one of our goals in this article to answer this question positively by showing that the null-tetrad approach can be extended to incorporate the influence of a plasma medium on the deflection angle and other optical quantities when static, asymptotically flat and spherically symmetric spacetimes are considered. 

Moreover, we will show that for spherically symmetric lenses it is also possible to write these optical quantities in terms of the different components of the  energy-momentum tensor and the parameters which characterize the plasma medium. These new expressions present a clear advantage with respect to the usual standard expressions based on derivatives and integration of different metric components, because in the last case it is not easy to see how the different components of $T_{\alpha\beta}$ contribute to the optical scalars. In particular, we showed in \cite{Gallo-lens-2011} that the deflection angle at first order can be written in terms of the energy-momentum tensor components as
\begin{equation}\label{angle-Gallointro}
\alpha=b\int_{-\lambda_{l}}^{\lambda_{ls}}\bigg[ 4\pi\bigg( \varrho+P_r \bigg)+\frac{b^2}{r^2}\bigg( \frac{3M}{r^3}-4\pi\varrho \bigg)\bigg]dy,
\end{equation}
where $b$ is the impact parameter, $\varrho$, $P_r$ and $M$ are the mass density, radial pressure and mass function respectively, $r=\sqrt{b^2+y^2}$, and $-\lambda_l$, and $\lambda_{ls}$ represent the value of the $y$-coordinate of the observer and source position respectively (the lens is assumed to be centered at $r=0$).

A crucial point in \cite{Gallo-lens-2011} that allowed us to
study the deflection angle and related quantities in terms of the curvature scalars was the use of the geodesic deviation equation which gives information of how a family of nearby null geodesics deviate from each other.

The first problem that arises when trying to extend that procedure to the study of light rays in a plasma media, is that photons in general do not follow null or timelike geodesics.  A detailed discussion and formulation for that problem using a geodesic deviation equation was recently presented in \cite{Schulze-Koops:2017tkc}.
However, we want to present here a more practical choice, which makes use of a 4-dimensional optical metric and where the computation of the different optical scalars follow from the study of the spatial projection of null geodesics in this \emph{optical} spacetime. 

An essential idea to
incorporate the plasma effect in this approach is the Gordon metric introduced by Gordon in 1923 \cite{W-1923}. As it is well known, this metric is an effective 4-dimensional Lorentzian metric whose null geodesics are in one-to-one  correspondence with timelike curves followed by light rays in a nondispersive media. 
However, even when this nice property does not remain valid for dispersive media,  we will see below that in the case of a static spacetime the spatial orbits of light rays derived from the study of null geodesics of a modified Gordon-like metric are the same as those obtained from the usual Hamiltonian approach. It will be precisely this result which will allow us to use the formalism developed by Gallo and Moreschi \cite{Gallo-lens-2011} to the description of the motion of photons even for dispersive media.

On the other hand, there is a very well known correspondence between propagation of photons in an homogeneous plasma and massive test particles in a pure gravity field \cite{Kulsrud-1992}. It is also our purpose to extend this correspondence to a more general kind of motion. In particular, we will establish for the first time a correspondence between the motion of light rays in a nonhomogeneous plasma and the orbits of massive charged test particles governed by an external electrical field in addition to gravity. The analogy will be possible due to the existence of a Riemannian metric (known as Jacobi metric) introduced by Gibbons \cite{Gibbons-Jacobi-static} and collaborators and exhaustively studied in later works \cite{Gibbons-Jac-Maup-1,Gibbons-Jac-Maup-2,Gibbons-Jac-Maup-3,Das:2016opi}. In this way, we will demonstrate that the Gibbons-Werner method originally conceived to study null geodesics, and extended by us to the case of plasma media, can also be used to study the nongeodesic motion of massive test particles that interact not only gravitationally with the background but also through external fields of nongravitational origin. 

This work is organized as follows. In Section \ref{section-orbit-equation} we show that the spatial orbits deduced from a Gordon-like metric associated with a static spacetime filled with a dispersive medium are exactly the same of those obtained from the usual Hamiltonian approach. In Section \ref{Gordon-like-null-tetrad} we use the same metric to incorporate the plasma effects in the optical scalars and the deflection angle through the null tetrad approach. In Section \ref{deflection-angle-energy-momentum} we compute the deflection angle and the optical scalars in spherically symmetric spacetimes for arbitrary distributions of matter and for any electronic charge density profile using the null tetrad approach and the Gauss-Bonnet method. In Section \ref{analogy-plasma-charged-particle} we show an analogy between the nongeodesic motion of charged massive particles and the one of photons propagating in a nonhomogeneous plasma, and apply these results to the study of the deflection angle of a massive charged test particle in a Reissner-Nordstr\"om spacetime. An Appendix with details and clarifications is also included.

\section{Spatial orbits associated to a Gordon-like metric}\label{section-orbit-equation}

Let us consider a light ray 
with 4-momentum $p_\alpha$ propagating in a general spacetime $(\mathcal{M},g_{\alpha\beta})$ and 
filled with a dispersive medium moving at velocity 
$u^\alpha$. In that spacetime, the photon frequency measured by an observer at rest with respect to the medium will be given by $\omega(x^\alpha)=p_\alpha u^\alpha$. 
The dispersive medium is completely 
characterized by its refractive 
index, a function of the 
coordinates and the frequency $\tilde{n}(x^\alpha,\omega(x^\alpha))$ that satisfies the following dispersion relationship \cite{book:75670},
\begin{equation}
    \tilde{n}^2=1-\frac{p_\alpha p^\alpha}{(p_\beta u^\beta)^2}.
\end{equation}
In that situation the Hamiltonian governing the dynamics of the photon is given by
\begin{equation}\label{eq:h1nue}
    H(x,p)=\frac{1}{2}(g^{\alpha\beta}+(\tilde{n}^2-1)u^\alpha u^\beta)p_\alpha p_\beta.
\end{equation}
and the light rays paths are obtained as solutions of Hamilton's equations,
\begin{equation}\label{ham-jac}
    \frac{dx^\alpha}{d\lambda}=\frac{\partial H}{\partial p_\alpha}, \ \ \ \ \frac{dp_\alpha}{d\lambda}=-\frac{\partial H}{\partial x^\alpha},
\end{equation}
with the constraint 
\begin{equation}
{H}(x,p)=0.\label{eq:consth}
\end{equation}
In 1923 Gordon was the first to introduce a symmetric tensor $\tilde{g}_{\alpha\beta}$ with respect to which it is possible to describe the propagation of photons in a  medium characterized by a refractive index $\tilde{n}(x^a,\omega)$ \cite{W-1923}. This tensor can be written as,
\begin{equation}\label{ham-jacz}
    \tilde{g}^{\alpha\beta}(x,p)=g^{\alpha\beta}+(\tilde{n}^2-1)u^\alpha u^\beta,
\end{equation}
with inverse $\tilde{g}_{\alpha\beta}$, defined as $\tilde{g}^{\alpha\beta}\tilde{g}_{\alpha\gamma}=\delta_\gamma^\beta$,
\begin{equation}\label{eq:gmetr}
    \tilde{g}_{\alpha \beta}(x,p)=g_{\alpha \beta}-\bigg(1-\frac{1}{\tilde{n}^2}\bigg)u_\alpha u_\beta.
\end{equation}
In terms of this tensor the Hamiltonian \eqref{eq:h1nue} reads,
\begin{equation}\label{H-gordon}
    \tilde{H}(x,p)=\frac{1}{2}\tilde{g}^{\alpha \beta}(x,p)p_\alpha p_\beta,
\end{equation}
In general this tensor is not a metric, because it also depends of the four-momentum $p_\alpha$ through $\omega$. From \eqref{H-gordon} and \eqref{ham-jacz} it follows that the equations of motion can be written as 
\begin{eqnarray}
    \frac{dx^\alpha}{d\lambda}&=&\tilde{g}^{\alpha\beta}p_\beta+\frac{\partial \tilde{n}^2}{\partial\omega}\omega^2 u^\alpha, \label{ham-jacx}\\ \frac{dp_\alpha}{d\lambda}&=&-\frac{1}{2}\tilde{g}^{\beta\gamma}_{\;\;\;\;,\alpha}p_\beta p_\gamma.\label{ham-jacp}
\end{eqnarray}
Note that if the medium described by the refractive index $\tilde{n}$ is nondispersive, that is 
\begin{equation}
    \frac{\partial \tilde{n}}{\partial \omega}=0,
\end{equation}
it follows that $\tilde{g}^{\alpha \beta}$ will depend only of the spacetime coordinates $x^\alpha$ and it becomes a Lorentzian metric \cite{book:75670}. Moreover, in such a case the second term of \eqref{ham-jacx} vanishes and hence the light rays becomes null geodesics of the metric $\tilde{g}_{\alpha\beta}$.
That is, the Gordon tensor reduces to a Lorentzian metric (known as the Gordon optical metric) and the motion of the light rays in the physical spacetime can be alternatively described as null geodesics of the $\tilde{g}_{\alpha\beta}$ given by \eqref{eq:gmetr}.

The situation is different for a dispersive media, in that case the second term of \eqref{ham-jacx} is not vanishing, and therefore the light rays do not follow null geodesics with respect to $\tilde{g}_{\alpha\beta}$. However, let us also consider a static spacetime with a timelike Killing vector $\xi^\alpha=\big(\frac{\partial}{\partial t}\big)^\alpha$ written as
\begin{equation}
    ds^2=g_{00}dt^2+g_{ij}dx^idx^j;
\end{equation}
filled with a dispersive medium which is also static, that is, the spatial components of the 4-velocity $u^\alpha$ of the medium with respect to the coordinate system $\{t,x^i\}$ are zero. The resulting tensor is therefore dependent on the spatial coordinates $x^i$, possibly of $N$ physical parameters $s_N$ characterizing the geometry and the 4-momentum $p_\alpha$ trough $\omega=p_\alpha u^\alpha=p_0 u^0$. The corresponding equations of motion for the light rays reduce to 
   \begin{eqnarray}
    \frac{dx^0}{d\lambda}&=&\tilde{g}^{00}p_0+\frac{\partial \tilde{n}^2}{\partial\omega}\omega^2 u^0, \label{ham-jacx2b0}\\
    \frac{dx^i}{d\lambda}&=&\tilde{g}^{ij}p_j, \label{ham-jacx2bi}\\
    \frac{dp_0}{d\lambda}&=&0\label{ham-jacp2b0},\\
    \frac{dp_i}{d\lambda}&=&-\frac{1}{2}\tilde{g}^{\beta\gamma}_{\;\;\;\;,i}p_\beta p_\gamma.\label{ham-jacp2bi}
\end{eqnarray}
together with the constraint \eqref{eq:consth}. 

Motivated by these equations of motion, we would like to construct a slightly modified Gordon metric depending only on the spatial coordinates (without any $p_\alpha$ dependence) and in such a way that even when their associated null geodesics will be not in one-to-one correspondence with the spacetime curves determined by the equations \eqref{ham-jacx2b0}-\eqref{ham-jacp2bi}, the spatial orbits will be the same. Note that in the static case, $\omega=p_0 u^0$ and therefore, the refractive index $\tilde{n}$ for a dispersive medium in general will have a dependence on $p_0$, in particular $\frac{\partial n}{\partial p_0}=\frac{\partial n}{\partial\omega}u^0\neq 0$. In order to construct a new metric that avoids this dependency we introduce the quantity $\hat\omega=\hat{p}_0u^0$, where $\hat{p}_0$ is a new parameter unrelated for the moment with the timelike component $p_0$ of four momentum $p_\alpha$ of the photon. 
In order to do that, let us consider a new metric tensor $\hat{g}_{\alpha\beta}(x)$ in the same way as in \eqref{eq:gmetr} but where the refractive index is replaced by a function only depending on the spatial coordinates, assuming the metric is static. This proxy function for the refractive index is given by $n=n(x^i)=\tilde{n}(x^i, \hat\omega(x^i))$, that is, we have replaced $\omega$ (which depends on $p_\alpha$) by $\hat\omega$ which does not share this dependence.
By replacing this expression for the refractive index into \eqref{eq:gmetr} we obtain the following Gordon-like optical metric,
\begin{equation}\label{eq:gmetr2}
    \hat{g}_{\alpha \beta}(x,s_N,\hat{p}_0)=g_{\alpha \beta}-\bigg(1-\frac{1}{n^2}\bigg)u_\alpha u_\beta,
\end{equation}
where we have made explicit the dependence of the resulting metric tensor $\hat{g}_{\alpha\beta}(x,s_N,\hat{p}_0)$ on the $N$ physical parameters $s_N$ describing the physical metric and also on the new parameter $s_{N+1}=\hat{p}_0$.
The null geodesics of $\hat{g}_{\alpha\beta}$ follow from 
\begin{equation}\label{eq:lade}
    \hat{H}=\frac{1}{2}\hat{g}^{\alpha\beta}(x,s_N,\hat{p}_0)p_\alpha p_\beta
    \end{equation}
    which is an homogeneous function of $p_\alpha$, together with the constraint $\hat{H}=0$ and are given by 
\begin{eqnarray}
    \frac{dx^0}{d\lambda}&=&\hat{g}^{00}(x,s_N,\hat{p}_0)p_0, \label{ham-jax2b0}\\
    \frac{dx^i}{d\lambda}&=&\hat{g}^{ij}(x,s_N,\hat{p}_0)p_j, \label{ham-jax2bi}\\
    \frac{dp_0}{d\lambda}&=&0\label{ham-jap2b0}\\
    \frac{dp_i}{d\lambda}&=&-\frac{1}{2}\hat{g}^{\beta\gamma}_{\;\;\;\;,i}(x,s_N,\hat{p}_0)p_\beta p_\gamma.\label{ham-jap2bi}
\end{eqnarray}
Now, let us assume that we solve these equations of motion by setting the new parameter $\hat{p}_0$ to take the same numerical value that $p_0$ (which must be a constant throughout the world-lines of the light rays because of \eqref{ham-jap2b0}). In that situation, we see that 
with the exception of the first of these expressions, all of them agree with equations \eqref{ham-jacx2b0}-\eqref{ham-jacp2bi}, and in particular the spatial orbit will be exactly the same. That is, the spatial orbit of the timelike curves followed by light rays in a dispersive medium in the physical metric $g_{\alpha\beta}$ are the same of those which follow from the study of the null geodesics of $\hat{g}_{\alpha\beta}$ satisfying Eqs.\eqref{ham-jax2b0}-\eqref{ham-jap2bi} after setting the parameter $\hat{p}_0$ to take the value $p_0$. As the approach to study optical scalars presented in \cite{Gallo-lens-2011} depends on integration on null geodesics, this result will be very useful in order to extend it to dispersive media.

Until now our considerations have been general. From here, we will concentrate on the situation where the dispersive media is given by a cold plasma.
That is, let us consider a static spacetime $(\mathcal{M},g_{\alpha\beta})$ filled with a cold nonmagnetized plasma described by the refractive index,
\begin{equation}\label{eq:nprime}
    \tilde{n}^2(x,\omega(x))=1-\frac{\omega_e^2(x)}{\omega^2(x)}.
\end{equation}
As before, in the previous expression $\omega(x)$ is the photon frequency as measured by a static observer and $\omega_e(x)$ is the electron plasma frequency,
\begin{equation}
    \omega_e^2(x)=\frac{4\pi e^2}{m_e}N(x),
\end{equation}
where $e$ and $m_e$ are the charge and mass of the electron respectively; $N(x)$ is the electron number density in the plasma.

In this kind of plasma, the Hamiltonian  governing the trajectory of photons given by \eqref{eq:h1nue} reduces to \cite{Perlick-book},
\begin{equation}\label{H-perlick}
    H(x,p)=\frac{1}{2} (g^{\alpha\beta}p_\alpha p_\beta + \omega_e^2(x)),
\end{equation}
where the path of light rays are obtained as solutions of the Hamilton's equations,
\begin{equation}\label{ham-jac}
    \frac{dx^\alpha}{d\lambda}=\frac{\partial H}{\partial p_\alpha}, \ \ \ \ \frac{dp_\alpha}{d\lambda}=-\frac{\partial H}{\partial x^\alpha},
\end{equation}
with the constraint $H(x,p)=0$.

If we consider a static and spherically symmetric spacetime
\begin{equation}\label{metrica-fisica}
    g_{\alpha\beta}dx^\alpha dx^\beta=A(r)dt^2-B(r)dr^2-C(r)d\Omega^2,
\end{equation}
where $d\Omega^2=d\vartheta^2+\sin^2\vartheta d\varphi^2$ is the metric of a unit sphere and we also consider a plasma density profile with the same symmetries, that is $\omega_e^2(x)=\omega_e^2(r)$, it is easy to show from \eqref{ham-jac} that the spatial orbits are described by the equation \cite{Perlick-book}
\begin{equation}\label{orbita-fisica}
    \bigg(\frac{d r}{d\varphi}\bigg)^2 = \frac{C(r)}{B(r)}\bigg( \frac{p_t^2}{p_\varphi^2} \frac{C(r)\tilde{n}^2(r)}{A(r)} -1 \bigg),
\end{equation}
with $p_t\equiv p_0$ and $p_\varphi$ result both constants of motion associated to the energy and angular momentum respectively.

As we discussed above, if the medium is dispersive, as in the case of a cold plasma, the solutions of Hamilton's equations associated with \eqref{eq:lade} do not agree with the ones associated with  \eqref{H-perlick} and then the Hamiltonian $\hat{H}$ does not describe the actual trajectory of a photon in the spacetime. Despite of this limitation, the spatial orbits coincide.

Let us see this particular case in more detail. If we assume that the plasma is static with respect to the observers following integral curves of the Killing vector field $\frac{\partial}{\partial t}$, we can take
\begin{equation}
    u^\alpha=\frac{\delta_0^\alpha}{\sqrt{A(r)}}.
\end{equation}
In this case the Gordon-like metric $\hat{g}_{\alpha \beta}$  is given by the expression,
\begin{equation}\label{Gordon-metric-spherical}
    \hat{g}_{\alpha \beta} dx^\alpha dx^\beta=\frac{A(r)}{{n}^2(r)}dt^2-B(r) dr^2-C(r)d\Omega^2,
\end{equation}
with
\begin{equation}
    {n}^2(r)=1-\frac{\omega^2_e(r)}{\hat\omega(r)}=1-\frac{\omega^2_e(r)A(r)}{\omega^2_\infty},
\end{equation}
where we have used that $\hat{\omega}(r)=\hat{p}_0u^0=\omega_\infty/\sqrt{A(r)}$, and we have identified $\hat{p}_0$ with the numerical value of which be later the photon frequency $\omega_\infty$ measured by an observer at the asymptotic region. As discussed above, this is only a parameter which takes the numerical value of $\omega_\infty$ and not should be thought a dynamical variable depending on $p_\alpha$.

Due to the spherical symmetry, we can take $\vartheta=\pi/2$ without loss of generality and from the Hamiltonian $\hat{H}(x,p)=\frac{1}{2}\hat{g}^{\alpha\beta}(x)p_\alpha p_\beta$ with the constraint $\hat{H}(x,p)=0$ we get that the null geodesics must satisfy,
\begin{equation}\label{constr2}
    \frac{p_t^2 n^2(r) }{A(r)}-\frac{p_r^2}{B(r)}-\frac{p_\varphi^2}{C(r)}=0;
\end{equation}
where again $p_t$ is a rename of $p_0$.
By using the Hamilton's equations associated to $\hat{H}$ we can see that
\begin{equation}
    \frac{dr}{d\lambda}=-\frac{p_r}{B(r)}, \ \ \ \ 
    \frac{d\varphi}{d\lambda}=-\frac{p_\varphi}{C(r)};
\end{equation}
and that $p_t$ and $p_\varphi$ are conserved quantities associated to the Killing vectors $\frac{\partial}{\partial t}$ and $\frac{\partial}{\partial\varphi}$ respectively.
Therefore the spatial orbital equation reads,
\begin{equation}
    \frac{d r}{d\varphi}=\frac{C(r)}{B(r)}\frac{p_r}{p_\varphi}.
\end{equation}
Finally, upon substituting for $p_r$ from \eqref{constr2} we get, 
\begin{equation}
    \bigg(\frac{d r}{d\varphi}\bigg)^2 = \frac{C(r)}{B(r)}\bigg( \frac{p_t^2}{p_\varphi^2} \frac{C(r)n^2(r)}{A(r)} -1 \bigg),
\end{equation}
which exactly matches with the spatial orbits \eqref{orbita-fisica} described by the physical metric \eqref{metrica-fisica} if we identify the respective constants of motion.

It is important to note that a necessary condition to obtain the equivalence for the spatial orbits  was to consider the dispersive medium to be at rest with respect to the static observers. In general, if we consider a 4-velocity with a nonvanishing spatial part ($u^i\neq 0$), in the equations of motion derived from the  Gordon-like metric will lack terms of the form $\frac{\partial n^2}{\partial\omega}\omega^2 u^i$ which otherwise are present in the true spatial orbits  
of the photons in the  presence of plasma determined by the Eq.\eqref{ham-jacx}.

\section{Gordon-like metric applied to the study of gravitational lenses using the null tetrad approach}\label{Gordon-like-null-tetrad}

In this section we argue that because the spatial orbits of light rays in a plasma can be found by studying null geodesics in the associated 4-dimensional Gordon-like metric, it is possible to apply the null tetrad approach with a gravitational lens surrounded by plasma. The null tetrad approach was used by Gallo and Moreschi in \cite{Gallo-lens-2011} to study the bending angle and optical scalars in pure gravity. 

\subsection{Review of the null tetrad approach in gravitational lensing}

Let us consider a past-null geodesic congruence starting at the observer position $O$ and ending at the source position $S$. The tangent vector to a fiducial null geodesic of this
congruence is given by $\ell=\frac{\partial}{\partial\lambda}$. The observer is assumed to be placed at $\lambda=0$ and the source at $\lambda=\lambda_s$. We complete a null tetrad $\{\ell^\alpha,n^\alpha,m^\alpha,\bar{m}^\alpha\}$ at the observer position where $m^\alpha$ and $\bar{m}^\alpha$ are orthogonal to the observer 4-velocity $u^\alpha$. Then we parallel transport the tetrad along the congruence up to the source position.

The deviation vector, with vanishing Lie derivative along the congruence, can be written as
\begin{equation}
    \zeta^\alpha=\zeta \bar{m}^\alpha+\bar{\zeta}m^\alpha+\zeta_\ell \ell^\alpha.
\end{equation}
It has been well discussed in the literature (see for example  \cite{Bartelmann:2010fz,Seitz:1994xf,Frittelli:2000bc}) that the geodesic deviation equation which describes the deviation vector along the congruence can be reduced to
\begin{equation}\label{dev-eq}
 \ell(\ell(\mathcal{X})) = 
 - Q \mathcal{X},
\end{equation}
where 
\begin{equation}
\mathcal{X} = 
\begin{pmatrix}
\varsigma  \\
\bar\varsigma
\end{pmatrix}, \ \ \
Q=
\begin{pmatrix}
\Phi_{00} & \Psi_0 \\
\bar{\Psi}_0 & \Phi_{00}
\end{pmatrix}
\end{equation}
and
\begin{equation}
\Phi_{00}=-\frac{1}{2}R_{\alpha\beta}\ell^\alpha\ell^\beta, \ \ \
\Psi_0=C_{\alpha\beta\gamma\delta}\ell^\alpha m^\beta \ell^\gamma m^\delta.
\end{equation}
Although equation \eqref{dev-eq} does not determine the evolution of the component $\zeta_\ell$, it is sufficient to obtain an expression for the optical scalars and deflection angle in the weak field approximation \cite{Gallo-lens-2011,Crisnejo-lens-2018}. To use this approach we only need to know how the deviation vector changes in the plane expanded by $m^\alpha$ and $\bar{m}^\alpha$.

Let us assume that a thin lens is placed at $\lambda=\lambda_l$ and let us define $\lambda_{l s}=\lambda_s-\lambda_l$.
As explained in \cite{Crisnejo-lens-2018}, in this approach the normalized convergence and shear at second order defined as,
\begin{equation}
    \tilde{\kappa}=\frac{\lambda_s}{\lambda_{ls}\lambda_l}\kappa, \ \ \ \tilde{\gamma}=\frac{\lambda_s}{\lambda_{ls}\lambda_l}\gamma,
\end{equation}
are given by the following expressions
\begin{equation}\label{convergence}
    \tilde{\kappa}=\tilde{\kappa}_{\text{\tiny $\Phi^{(1)}$}}+\tilde{\kappa}_{\text{\tiny $\Phi^{(2)}$}}+\tilde{\kappa}_{\text{\tiny $\delta$}\text{\tiny $\Phi$}}+\tilde{\kappa}_{\text{\tiny $\Phi$}\text{\tiny $\Phi$}}+\tilde{\kappa}_{\text{\tiny $\Psi$}\text{\tiny $\Psi$}},
\end{equation}
where 
\begin{eqnarray}
    \tilde{\kappa}_{\text{\tiny $\Phi^{(i)}$}}&=&\int_0^{\lambda_s}\Phi_{00}^{(i)}d\lambda, \ \ \ \ i = 1,2, \\
    \tilde{\kappa}_{\text{\tiny $\delta$}\text{\tiny $\Phi$}}&=& \int_0^{\lambda_s}\delta x^{(1)\alpha}(\lambda)\frac{\partial \Phi_{00}^{(1)}}{ \partial x^\alpha}\bigg|_{x^(0)(\lambda)}d\lambda, \\
    \tilde{\kappa}_{\text{\tiny $\Phi$}\text{\tiny $\Phi$}}&=& -\frac{1}{\lambda_l \lambda_{ls}}\int_0^{\lambda_s}\int_0^\lambda \lambda^\prime (\lambda_s-\lambda)(\lambda-\lambda^\prime)\\ \nonumber
&\times& \Phi_{00}^{(1)}(\lambda^\prime)\Phi_{00}^{(1)}(\lambda)d\lambda^\prime d\lambda, \\
\tilde{\kappa}_{\text{\tiny $\Psi$}\text{\tiny $\Psi$}}&=& -\frac{1}{\lambda_l \lambda_{ls}}\int_0^{\lambda_s}\int_0^\lambda \lambda^\prime (\lambda_s-\lambda)(\lambda-\lambda^\prime)\\ \nonumber
&\times& \Psi_{0}^{(1)}(\lambda^\prime)\Psi_{0}^{(1)}(\lambda)d\lambda^\prime d\lambda,
\end{eqnarray}
and
\begin{equation}\label{shear}
    \tilde{\gamma}=\tilde{\gamma}_{\text{\tiny $\Psi^{(1)}$}}+\tilde{\gamma}_{\text{\tiny $\Psi^{(2)}$}}+\tilde{\gamma}_{\text{\tiny $\delta$}\text{\tiny $\Psi$}}+\tilde{\gamma}_{\text{\tiny $\Phi$}\text{\tiny $\Psi$}}+\tilde{\gamma}_{\text{\tiny $\Psi$}\text{\tiny $\Phi$}},
\end{equation}
with
\begin{eqnarray}
    \tilde{\gamma}_{\text{\tiny $\Psi^{(i)}$}}&=&\int_0^{\lambda_s}\Psi_{0}^{(i)}d\lambda, \ \ \ \ i = 1,2, \\
    \tilde{\gamma}_{\text{\tiny $\delta$}\text{\tiny $\Psi$}}&=& \int_0^{\lambda_s}\delta x^{(1)\alpha}(\lambda)\frac{\partial \Psi_{0}^{(1)}}{ \partial x^\alpha}\bigg|_{x^(0)(\lambda)}d\lambda, \\
    \tilde{\gamma}_{\text{\tiny $\Phi$}\text{\tiny $\Psi$}}&=& -\frac{1}{\lambda_l \lambda_{ls}}\int_0^{\lambda_s}\int_0^\lambda \lambda^\prime (\lambda_s-\lambda)(\lambda-\lambda^\prime)\\ \nonumber
&\times& \Phi_{00}^{(1)}(\lambda^\prime)\Psi_{0}^{(1)}(\lambda)d\lambda^\prime d\lambda, \\
\tilde{\gamma}_{\text{\tiny $\Psi$}\text{\tiny $\Phi$}}&=& -\frac{1}{\lambda_l \lambda_{ls}}\int_0^{\lambda_s}\int_0^\lambda \lambda^\prime (\lambda_s-\lambda)(\lambda-\lambda^\prime)\\ \nonumber
&\times& \Psi_{0}^{(1)}(\lambda^\prime)\Phi_{00}^{(1)}(\lambda)d\lambda^\prime d\lambda.
\end{eqnarray}
The notation has to be read as follows. The terms $\tilde{\kappa}_{\text{\tiny $\Phi^{(i)}$}}$ indicate the contribution to the convergence due to the scalar $\Phi_{00}$ at $i^{th}$ order, while the term $\tilde{\kappa}_{\text{\tiny $\delta$}\text{\tiny $\Phi$}}$ indicates that a first order correction to the photon path $\delta x^{(1)\alpha}$ is included via the term  
$\delta x^{(1)\alpha} \partial_\alpha \Phi_{00}^{(1)}$. The quantity $\Phi_{00}^{(1)}$ is the $\Phi_{00}$ scalar at the same order. The last two terms in \eqref{convergence} indicate the contribution by $\Phi_{00}^{(1)}(\lambda)\Phi_{00}^{(1)}(\lambda^{\prime})$ and $\Psi_{0}^{(1)}(\lambda)\Psi_{0}^{(1)}(\lambda^{\prime})$ respectively. Similar interpretation should be given for each term in \eqref{shear}. In addition in \cite{Crisnejo-lens-2018} we obtained an expression for the rotation optical scalar which is zero for spherically symmetric spacetimes (for more details see \cite{Crisnejo-lens-2018}). Finally, the deflection angle can be expressed in terms of $\tilde{\kappa}$ and $\tilde{\gamma}$ as,
\begin{equation}\label{alpha-kappa-gamma}
    \alpha=b(\tilde{\kappa}+\tilde{\gamma}),
\end{equation}
where $b$ is the impact parameter.

The advantage of this approach is that we can study separately the contribution to the optical scalars and deflection angle given by the energy-momentum distribution (through the Ricci tensor and Einstein equations) and by the Weyl curvature (through the Weyl tensor).

As we have shown in the previous section, light rays follow exactly the same spatial orbits both in the physical spacetime (whose dynamics is determined by the Hamiltonian \eqref{H-perlick}) and in the 4-dimensional optical spacetime (following null geodesics of \eqref{Gordon-metric-spherical}). Therefore, since the $t=$ constant spacelike sections of both spacetimes have the same spatial metric, in both spaces the angles are defined in the same way. In particular, it also applies to the deflection angle of light rays, which implies that we can use the machinery originally adapted only to cases of pure gravity to study deflections of light rays in a gravitational environment where a dispersive medium is present.

\subsection{Gravitational lenses surrounded by an homogeneous plasma}
In this subsection we will use two simple examples to demonstrate the use of the previous formalism to calculate optical scalars and deflection angle in an homogeneous plasma medium $(\omega_e=\text{constant})$. 

Let us first examine the  Schwarzschild lens model at linear  order, and after that a situation a little more complicated: the \emph{parametrized-post-Newtonian} (PPN) lens model, at second order.

\subsubsection{Schwarzschild lens model}

Let us calculate the first order contributions to the deflection angle and optical scalars for a static spherically symmetric lens described by the Schwarzschild metric (in isotropic coordinates),
\begin{equation}\label{Schw-metric}
    ds^2=\frac{(1-\frac{m}{2r})^2}{(1+\frac{m}{2r})^2} dt^2-\bigg(1+\frac{m}{2r}\bigg)^4 d\vec{x}^2,
\end{equation}
where $d\vec{x}^2=dx^2+dy^2+dz^2$ and $r=\sqrt{x^2+y^2+z^2}$. We will consider the case in which the lens is surrounded by an homogeneous plasma ($\omega_e$=constant) with refractive index
\begin{equation}
    n^2(r)=1-\frac{\omega^2_e}{\omega^2(r)}=1-\frac{\omega_e^2}{\omega_\infty^2}\frac{(1-\frac{m}{2r})^2}{(1+\frac{m}{2r})^2},
\end{equation}
where the gravitational redshift has already been taken into account, that is $\omega(r)=\omega_\infty/\sqrt{A(r)}$ with $\omega_\infty$ the photon frequency as measured by an asymptotic observer which can be related to the frequency $\omega_o$ measured by an observer which is placed at finite distance $r_o$ from the lens by $\omega_\infty=\omega_o\sqrt{A(r_o)}$. In the rest of this work, we always assume that the observer is at great distance from the lens, and therefore we can take $\omega_\infty\approx \omega_o$. 

The Gordon-like metric associated to \eqref{Schw-metric} is given by
\begin{equation}\label{Gordon-Schw}
    d\hat{s}^2=\frac{1}{n^2(r)}\frac{(1-\frac{m}{2r})^2}{(1+\frac{m}{2r})^2} dt^2-\bigg(1+\frac{m}{2r}\bigg)^4 d\vec{x}^2.
\end{equation}
By doing the change of variable,
\begin{equation}
    \tilde{t}=\frac{t}{\sqrt{1-\frac{\omega_e^2}{\omega_\infty^2}}},
\end{equation}
and expanding at first order in the mass parameter, we obtain the following expression for the Gordon-like metric,
\begin{equation}
    d\hat{s}^2\approx\bigg( 1-\frac{2m}{r n_o^2} \bigg)d\tilde{t}^2-\bigg( 1+\frac{2m}{r} \bigg)d\vec{x}^2,
\end{equation}
where we have defined,
\begin{equation}
    n_o=\sqrt{1-\frac{\omega_e^2}{\omega_\infty^2}}.
\end{equation}
In order to compute the bending angle and the optical scalars using the formalism developed in this section, we have to integrate the curvature scalars $\Phi_{00}$ and $\Psi_0$  along the actual null curve followed by a photon from the source to the receiver. In the weak gravitational lensing regime the actual path can be thought as a null geodesic in the flat background plus higher order corrections. Since the curvature scalars are already linear order quantities, in order to compute the optical scalars and the  deflection angle at this order it is enough to approximate the actual path by the null geodesic on the background. This is known as the Born approximation.

We choose a Cartesian coordinate system with respect to which this curve propagates in the $y$ negative direction and we will perform the integral toward the past from the observer to the source position (The observer is assumed to be at the asymptotic region). Then the tetrad adapted to this curve in the background can be selected as (see \cite{Gallo-lens-2011} for more details),

\begin{equation}\label{eq:tetradflat}
\begin{split}
l^a=&(-1,0,1,0),\ \
m^a=\frac{1}{\sqrt{2}}(0,i,0,1),\\
\bar{m}^a=& \frac{1}{\sqrt{2}}(0,-i,0,1),\ \
n^a=\frac{1}{2}(-1,0,-1,0).
\end{split}
\end{equation}
It is convenient to introduce coordinates $b$ and $\vartheta$ representing the impact parameter and polar angle measured from $z$ related to Cartesian coordinates by,
\begin{equation}
\begin{aligned}
    z=&b\cos(\vartheta)=0,\\
    x=&b\sin(\vartheta)=b,
\end{aligned}
\end{equation}
where in the last equality we used the spherical symmetry which allows us to work in the plane $\vartheta=\pi/2$ without loss of generality.

As in  previous works \cite{Gallo-lens-2011,Crisnejo-lens-2018}, we choose the origin of the coordinate system in the lens's position $\lambda_l$ and parametrize the geodesic by
\begin{equation}\label{eq:parametrizaci}
\begin{aligned}
    (x(\lambda),y(\lambda),z(\lambda))=&(x,y-\lambda_{l},z) \\
    =&(b,y-\lambda_{l},0).
\end{aligned}    
\end{equation}

By computing $\Phi_{00}$ and $\Psi_0$ we get
\begin{equation}
\begin{aligned}
    \Phi_{00}=\frac{m}{2}\frac{b^2-2\lambda^2+4\lambda \lambda_l-2 \lambda_l^2}{(b^2+(\lambda-\lambda_l)^{2})^{5/2}} \bigg(1-\frac{1}{n_o^2}\bigg),
\end{aligned}
\end{equation}
\begin{equation}
\begin{aligned}
    \Psi_{0}=-\frac{3m}{2}\frac{b^2}{(b^2+(\lambda-\lambda_l)^{2})^{5/2}}\bigg(1+\frac{1}{n_o^2}\bigg).
\end{aligned}
\end{equation}
In particular, we can see that for the vacuum case $(n_o^2=1)$ we get $\Phi_{00}=0$ and consequently the convergence is zero.

By using the relations \eqref{convergence} and \eqref{shear} at first order one obtains for the optical scalars,
\begin{equation}
\begin{aligned}
\kappa =& 0, \\
\gamma =& \frac{\lambda_l \lambda_{ls}}{\lambda_s} \frac{2m}{b^2} \bigg(1+\frac{1}{1-\omega_{e}^2/\omega_{\infty}^2}\bigg).
\end{aligned}
\end{equation}
Notice that even when the Gordon-like metric does not satisfy the Einstein's field equations, the convergence coincides with its value in the vacuum case which is zero. That is, at first order the homogeneous plasma has not influence in the convergence.

Finally, from \eqref{alpha-kappa-gamma} we obtain the deflection angle,
\begin{equation}
    \alpha= \frac{2m}{b} \bigg(1+\frac{1}{1-\omega_{e}^2/\omega_{\infty}^2}\bigg),
\end{equation}
which is in complete agreement with the known expression found using a standard coordinate-dependent method\cite{BisnovatyiKogan:2010ar}.

\subsubsection{Parametrized-post-Newtonian (PPN) lens model}

In order to compute the optical scalars and deflection angle at second order with this approach we will consider a more general lens model described by the \emph{parametrized-post-Newtonian} (PPN) metric whose line element is given by,
\begin{equation}
    ds=\bigg( 1-\frac{2m}{r}+\frac{2\beta m^2}{r^2} \bigg)dt^2-\bigg( 1+\frac{2\mu m}{r}+\frac{3\nu m^2}{2r^2} \bigg)d\vec{x}^2,
\end{equation}

By making the change of variable $\tilde{t}=\frac{t}{n_{o}}$, the Gordon-like metric associated with the PPN metric is given by,
\begin{equation}
\begin{aligned}
    d\hat{s}^2=&\bigg( 1-\frac{2m}{n_{o}^2 r}+\frac{2m^2}{n_{o}^4 r^2}(\beta n_{o}^2+2(1-n_o^{2})) \bigg) d\tilde{t}^2 \\ &-\bigg( 1+\frac{2\mu m}{r}+\frac{3\nu m^2}{2r^2} \bigg)d\vec{x}^2.
\end{aligned}
\end{equation}
In order to calculate the deflection angle we need to make a parallel transport of  the null tetrad \eqref{eq:tetradflat} at first order along the vector $\ell^a$. Then, we obtain
\begin{equation}
 \ell^{t} = -1 + \bigg(  \frac{1}{\sqrt{b^{2}+\lambda_{l}^{2}}} -\frac{2}{\sqrt{b^{2}+(\lambda-\lambda_{l})^{2}}} \bigg) \frac{m}{n_{o}^2}
 \end{equation}
 \begin{equation}
 \ell^{x} = \frac{(1+n_{o}^2\mu )}{b} \bigg( \frac{\lambda_{l}-\lambda}{\sqrt{b^{2}+(\lambda-\lambda_{l})^{2}}} 
	  - \frac{\lambda_{l}}{\sqrt{b^{2}+\lambda_{l}^{2}}} \bigg) \frac{m}{n_{o}^2},
 \end{equation}
 \begin{equation}
 \ell^{y} = 1 - \bigg( \frac{n_{o}^2\mu-1}{\sqrt{b^{2}+(\lambda-\lambda_{l})^{2}}} + \frac{1}{\sqrt{b^{2}+\lambda_{l}^{2}}} \bigg) \frac{m}{n_{o}^2},
\end{equation}
\begin{equation}
 \ell^{z} = 0;
\end{equation}

\begin{equation}
m^{t} = \frac{i}{b\sqrt{2}} \bigg(  \frac{\lambda-\lambda_{l} }{\sqrt{b^{2}+(\lambda-\lambda_{l})^{2}} } 
        + \frac{\lambda_{l} }{\sqrt{b^{2}+\lambda_{l}^{2}}} \bigg) \frac{m}{n_{o}^2},
\end{equation}
\begin{equation}
 m^{x} = \frac{i}{\sqrt{2}} -  \frac{i \, \mu}{\sqrt{2}} \frac{m}{\sqrt{b^{2}+(\lambda-\lambda_{l})^{2}}},
\end{equation}
\begin{equation}
 m^{y} = \frac{\mu}{\sqrt{2}} \frac{i}{ J} \bigg(  \frac{\lambda-\lambda_{l} }{\sqrt{b^{2}+(\lambda-\lambda_{l})^{2}} } 
        + \frac{\lambda_{l} }{\sqrt{b^{2}+\lambda_{l}^{2}}} \bigg) m,
\end{equation}
\begin{equation}
 m^{z} = \frac{1}{\sqrt{2}} - \frac{\mu}{\sqrt{2}} \frac{m}{\sqrt{b^{2}+(\lambda-\lambda_{l})^{2} }};     
\end{equation}
The correction to the background null geodesic, which follows from the integration of the $\ell^a$
components at first order is
\begin{equation}
\begin{aligned}
 \delta x^{t} =& \bigg( \frac{\lambda}{\sqrt{b^{2}+\lambda_{l}^{2}}} - 2\;\text{arcsinh}(\frac{\lambda-\lambda_{l}}{b}) \\
 &- 2\;\text{arcsinh}(\frac{\lambda_{l}}{b}) \bigg) \frac{m}{n_{o}^2}, 
 \end{aligned}
 \end{equation}
 \begin{equation}
 \begin{aligned}
 \delta x^{x} =& \frac{(n_{o}^2\mu + 1)}{b} \bigg( \sqrt{b^{2}+\lambda_{l}^{2}}-\sqrt{b^{2}+(\lambda-\lambda_{l})^{2}} \\
 &- \frac{\lambda\lambda_{l}}{\sqrt{b^{2}+\lambda_{l}^{2}}} \bigg) \frac{m}{n_{o}^2},
\end{aligned} 
 \end{equation}
 \begin{equation}
 \begin{aligned}
 \delta x^{y} =& \bigg[ (1-n_{o}^2\mu) \bigg(\text{arcsinh}(\frac{\lambda-\lambda_{l}}{b}) + \text{arcsinh}(\frac{\lambda_{l}}{b}) \\
 &- \frac{\lambda}{\sqrt{b^{2}+\lambda_{l}^{2}}} \bigg) - \frac{\mu\lambda}{\sqrt{b^{2}+\lambda_{l}^{2}}}\bigg] \frac{m}{n_{o}^2},
\end{aligned} 
\end{equation}
\begin{equation}
 \delta x^{z} = 0.
\end{equation}

\begin{table}[h]
\caption{\label{Table1}
Convergence. The computation of each terms in \eqref{convergence} for the PPN metric
}
\begin{tabular}{ l @{\qquad} c }
\toprule
\textrm{}&
\textrm{PPN}\\
\colrule
$\tilde{\kappa}_{\text{\tiny $\Phi^{(1)}$}}$ & 0\\
$\tilde{\kappa}_{\text{\tiny $\Phi^{(2)}$}}$ & $\frac{\pi m^2}{16 n_{o}^2 b^3}(\frac{3}{n_{o}^2}-16-6\nu n_{o}^2+2\mu+9n_{o}^2\mu^2+8 \beta)$ \\
$\tilde{\kappa}_{\text{\tiny $\delta$}\text{\tiny $\Phi$}}$ & $-\frac{\pi m^2}{8n_{o}^4 b^3}(-2+n_{o}^2\mu+n_{o}^4\mu^{2})$\\
$\tilde{\kappa}_{\text{\tiny $\Phi$}\text{\tiny $\Phi$}}$ & $\frac{\pi m^2}{32n_{o}^4 b^3}(1-2n_{o}^2\mu+n_{o}^4\mu^{2}) $\\
$\tilde{\kappa}_{\text{\tiny $\Psi$}\text{\tiny $\Psi$}}$ & $-\frac{15\pi m^2}{32n_{o}^4 b^3} (1+2n_{o}^2 \mu+n_{o}^4\mu^2)$\\
\botrule
\end{tabular}
\end{table}

\begin{table}[h]
\caption{\label{Table2}
Shear. The computation of each terms in \eqref{shear} for the PPN metric
}
\begin{tabular}{ l @{\qquad} c }
\toprule
\textrm{}&
\textrm{PPN}\\
\colrule
$\tilde{\gamma}_{\text{\tiny $\Psi^{(1)}$}}$ & $\frac{2m}{b^2}(\mu+\frac{1}{n_{o}^2})$\\
$\tilde{\gamma}_{\text{\tiny $\Psi^{(2)}$}}$ & $\frac{\pi m^2}{16 n_{o}^2 b^3}(48-\frac{3}{n_{o}^2}-24\beta+18 n_{o}^2\nu +6\mu-39 n_{o}^2\mu^2)$ \\
$\tilde{\gamma}_{\text{\tiny $\delta$}\text{\tiny $\Psi$}}$ & $\frac{3\pi m^2}{8n_{o}^4 b^3}(2+7n_{o}^2\mu+5 n_{o}^4 \mu^2)$\\
$\tilde{\gamma}_{\text{\tiny $\Phi$}\text{\tiny $\Psi$}}$ & $\frac{9\pi m^2}{32n_{o}^4 b^3}(n_{o}^4\mu^2 -1) $\\
$\tilde{\gamma}_{\text{\tiny $\Psi$}\text{\tiny $\Phi$}}$ & $\frac{9\pi m^2}{32n_{o}^4 b^3}(n_{o}^4\mu^2 -1) $\\
\botrule
\end{tabular}
\end{table}
The corresponding curvature scalars are given by
\begin{eqnarray}
    \Phi_{00}&=&\Phi^{(1)}_{00}+\Phi^{(2)}_{00},\\
    \Psi_{0}&=&\Psi^{(1)}_{0}+\Psi^{(2)}_{0}
\end{eqnarray}
where the expressions for the linear terms are given by
\begin{equation}
\begin{aligned}
    \Phi_{00}^{(1)} =\frac{m}{2}\frac{b^2-2\lambda^2+4\lambda \lambda_l-2 \lambda_l^2}{(b^2+(\lambda-\lambda_l)^{2})^{5/2}} \bigg(\mu-\frac{1}{n_o^2}\bigg),
\end{aligned}
\end{equation}
\begin{equation}
\begin{aligned}
    \Psi_{0}^{(1)} =-\frac{3m}{2}\frac{b^2}{(b^2+(\lambda-\lambda_l)^{2})^{5/2}}\bigg(\mu+\frac{1}{n_o^2}\bigg).
\end{aligned}
\end{equation}

The second order corrections terms are more complicated,
\begin{widetext}
\begin{equation}
 \begin{split}
   \Psi^{(2)}_0=& \frac{m^2}{[b^2+(\lambda-\lambda_l)^2]^{5/2}(\omega_\infty^2-\omega^2_e)^2}\times\bigg\{-\frac{3[\mu(\omega_\infty^2-\omega^2_e)+\omega_\infty^2]^2(2b^2+\lambda^2_l+\lambda\lambda_l)}{\sqrt{b^2+\lambda^2_l}}\\
    &+\frac{(\omega_\infty^2-\omega^2_e)}{2[b^2+(\lambda-\lambda_l)^2]^{1/2}}\bigg[b^2[2\omega_\infty^2(4\beta+\mu)+(\omega_\infty^2-\omega^2_e)(15\mu^2-6\nu)]-6(\omega_\infty^2-\omega^2_e)(\lambda-\lambda_l)^2\mu^2-12\omega_\infty^2(\lambda-\lambda_l)^2\mu\\
    &-\frac{13\omega_\infty^2(\omega_\infty^2-2\omega^2_e)b^2+6\omega_\infty^4(\lambda-\lambda_l)^2}{\omega^2_\infty-\omega^2_e}\bigg]\bigg\},
    \end{split}
\end{equation}

\begin{equation}
 \begin{split}
    \Phi^{(2)}_{00}=& \frac{m^2}{2(\omega_\infty^2-\omega_e^2)} \bigg\{ \frac{1}{(b^2+(\lambda-\lambda_l)^2)^3} \bigg[ b^2 \bigg(4\beta\omega_\infty^2-3\mu^2(\omega_\infty^2-\omega_e^2)+4\mu\omega_\infty^2\bigg) - \frac{b^2\omega_\infty^2}{\omega_\infty^2-\omega_e^2} (5\omega_\infty^2-8\omega_e^2)\\
    &+(\lambda-\lambda_l)^2 \bigg( 18\mu^2(\omega_\infty^2-\omega_e^2)-4\beta \omega_\infty^2-10\mu \omega_\infty^2-6\nu(\omega_\infty^2-\omega_e^2)+2\omega_\infty^2 \frac{\omega_\infty^2-4\omega_e^2}{\omega_\infty^2-\omega_e^2}\bigg)\bigg]\\
    &+\frac{1}{\sqrt{b^2+\lambda_l^2}(b^2+(\lambda-\lambda_l)^2)^{7/2}} \bigg[2b^4\bigg( \mu^2(\omega_\infty^2-\omega_e^2)-2\mu\omega_\infty^2\bigg)+2b^4 \frac{\omega_\infty^4}{\omega_\infty^2-\omega_e^2}+8\mu\omega_\infty^2(\lambda-\lambda_l)^4\\
    &+2b^2(\lambda-\lambda_l)\bigg(2\mu\omega_\infty^2(\lambda-\lambda_l)
    -\mu^2(\lambda-4\lambda_l)(\omega_\infty^2-\omega_e^2)\bigg)-2b^2(\lambda-\lambda_l)(\lambda+2\lambda_l)\frac{\omega_\infty^4}{\omega_\infty^2-\omega_e^2}\\
    &-2\mu^2 (\omega_\infty^2-\omega_e^2)(2\lambda-5\lambda_l)(\lambda-\lambda_l)^3
    -2(2\lambda+\lambda_l)(\lambda-\lambda_l)^3\frac{\omega_\infty^4}{\omega_\infty^2-\omega_e^2}\bigg]
    \bigg\}.
    \end{split}
\end{equation}
\end{widetext}
By replacing these expressions into Eqs.\eqref{convergence} and \eqref{shear} and taking the limits $\lambda_l\to\infty$ and $\lambda_{ls}\to\infty$ we obtain,
\begin{equation}
    \tilde{\kappa}=\frac{\pi m^2}{8n_{o}^2 b^3}(-8+4\beta-3 n_{o}^2\nu-8\mu) ,
\end{equation}
\begin{equation}
    \tilde{\gamma}=\frac{2m}{b^2}\bigg(\mu+\frac{1}{n_{o}^2}\bigg)+\frac{3}{8}\frac{\pi m^2}{n_{o}^2 b^3}(8-4\beta+3n_{o}^2\nu+8\mu).
\end{equation}
These expressions generalize the results recently derived by us in \cite{Crisnejo-lens-2018}. 

In the Tables \ref{Table1} and \ref{Table2} we can see the contribution of each term to the convergence and shear, respectively. Except for the term $\tilde{\kappa}_{\text{\tiny $\Phi^{(1)}$}}$, which remains equal to zero as in the pure gravity case, each of them are modified by the presence of the plasma. In particular, if we only consider first order corrections in the mass parameter, the presence of the plasma will contribute to the shear but not to the convergence.

Finally, from the relation \eqref{alpha-kappa-gamma} we obtain the deflection angle
\begin{equation}
    \alpha=\frac{2m}{b}\bigg(\mu+\frac{1}{n_{o}^2}\bigg)+\frac{\pi m^2}{ b^2}\bigg( \frac{2-\beta+2\mu}{n_{o}^2}+\frac{3}{4}\nu \bigg),
\end{equation}
recovering the result obtained previously by us in \cite{Crisnejo:2018ppm} using the Gauss-Bonnet theorem. 

\section{Deflection angle and optical scalars in terms of energy-momentum distributions for lenses surrounded by plasma}\label{deflection-angle-energy-momentum}
In \cite{Gallo-lens-2011} Gallo and Moreschi have shown  that for spherically symmetric spacetimes, expressions for the deflection angle and optical scalars can be obtained in terms of the energy-momentum distribution. Particular cases of these expressions were recently reobtained by De Leon and Vega using the Gauss-Bonnet theorem \cite{deLeon:2019qnp}. Deriving these expressions require use of a null tetrad adapted to the geometry of the matter distribution. In addition, we must find the transformation between this tetrad and one adapted to the motion of photons at first order (\cite{Gallo-lens-2011}). 

As we mentioned in the Introduction, to have expressions for the deflection angle in terms of the energy-momentum tensor instead of the metric components have several advantages. We will see how to extend this approach to the case of lenses surrounded by a cold nonmagnetized plasma. Finally, we will compare these results with those obtained by using the alternative geometrical method based on the Gauss-Bonnet theorem. 

\subsection{Using the null tetrad approach}\label{gallo-mor}
Let $(\mathcal{M},g_{\alpha\beta})$ be a static, spherically symmetric and asymptotically flat spacetime with a line element given by
\begin{equation}\label{metric-spherically-symm}
 ds^2 = A(r) dt^2 - B(r) dr^2 - r^2(d\theta^2 + \sin^2 \theta d\varphi^2),
\end{equation}
with
\begin{equation}
 A(r) = e^{2 \Phi(r)}, \ \ B(r) = \frac{1}{1 - \frac{2 M(r)}{r} }.
\end{equation}
A more general  energy-momentum distribution compatible with spherical 
symmetry is described by an energy-momentum tensor given by
\begin{equation}
 T_{tt} =  \varrho e^{2 \Phi(r)};
\end{equation}
\begin{equation}
 T_{rr} =    \frac{P_r}{\left( 1 - \frac{2 M(r)}{r} \right) } ;
\end{equation}
\begin{equation}
 T_{\theta \theta} =    P_t \, r^2 ;
\end{equation}
\begin{equation}
 T_{\varphi \varphi} =    P_t \,  r^2 \sin(\theta)^2 ;
\end{equation}
where we have introduced the notion of radial component  $P_r$ and
tangential component  $P_t$.

The Einstein field equations
\begin{equation}
G_{\alpha\beta} = -8\pi T_{\alpha\beta} ,
\end{equation} 
in terms of the previous variables are \cite{Gallo-lens-2011}
\begin{equation}\label{eq:rho}
 \frac{dM}{dr} = 4\pi r^2 \varrho ,
\end{equation}
\begin{equation}\label{eq:mgrt}
r^2\frac{d\Phi}{dr} = \frac{M + 4\pi r^3 P_r}{1 - \frac{2 M(r)}{r} } ,
\end{equation}
\begin{equation}
\begin{split}\label{eq:pt}
 r^3 &\left(\frac{d^2\Phi}{dr^2}+(\frac{d\Phi}{dr})^2\right)( 1-\frac{2 M}{r})\\
&+ r^2 \frac{d\Phi}{dr} (1 - \frac{M}{r}-\frac{dM}{dr})\\
&- r \frac{dM}{dr} + M 
= 8\pi r^3 P_t  .
\end{split}
\end{equation}
The conservation equation is 
\begin{equation}\label{eq:consrt}
 \frac{dP_r}{dr} = -(\varrho + P_r ) \frac{d\Phi}{dr} 
- \frac{2}{r} (P_r - P_t)  .
\end{equation}

\subsubsection{Gordon-like metric}

In Section \ref{section-orbit-equation} we have proved that spatial orbits followed by photons in a cold nonmagnetized plasma are the same for both the physical metric $g_{\alpha\beta}$ and an effective optical metric $\hat{g}_{\alpha\beta}$ given by \eqref{Gordon-metric-spherical},
where for a spherically symmetric cold nonmagnetized plasma the index of refraction is given by
\begin{equation}\label{eq:indrefp}
    n^2(r)=1-\frac{\omega_e^2(r)}{\omega_\infty^2}A(r).
\end{equation}
As before, $\omega_\infty$ is the photon frequency measured by an asymptotic observer. 

As we are interested in small deviation from the flat background it is enough to make the following approximations to the metric components,
\begin{eqnarray}
A(r) &\approx& 1+2\Phi(r),\\
B(r) &\approx& 1+\frac{2M(r)}{r} =: \tilde{B}(r).\label{coef-B}
\end{eqnarray}
In addition, let us consider electron plasma frequencies of the form,
\begin{equation}\label{omega-plasma-comp-1}
    \omega_e^2(r) \approx \omega_{e0}^2+K_{e}N_{1}(r),
\end{equation}
where $K_e=\frac{4\pi e^2}{m_e}$, being $e$ and $m_e$ the charge and mass of electron, respectively. In this case we are assuming that $\omega_{e0}^2=\text{constant}$ and
\begin{equation}\label{omega-plasma-comp-2}
\frac{K_e N_{1}(r)}{\omega_{e0}^2}\ll 1, \ \ \ \lim_{r\to\infty} N_1(r)=0.
\end{equation}
It is convenient to introduce the definition
\begin{equation}
    \tilde{n}_{o}=\sqrt{1-\frac{\omega_{e0}^2}{\omega_{\infty}^2}},
\end{equation}
and make the change of variable
\begin{equation}
    \tilde{t}=\frac{t}{\tilde{n}_{o}}.
\end{equation}
In this way the Gordon-like metric can be expressed as follows,
\begin{equation}\label{effective-metric-2}
 d\hat{s}^2 = \tilde{A}(r) d\tilde{t}^2 - \tilde{B}(r) dr^2 - r^2(d\theta^2 + \sin^2 \theta d\varphi^2),
\end{equation}
where
\begin{equation}\label{coef-A}
    \tilde{A}(r)=1+\frac{2\Phi(r)}{\tilde{n}_{o}^{2}}+\frac{K_e N_1(r)}{\omega_{\infty}^2 \tilde{n}_{o}^2},
\end{equation}
and $\tilde{B}(r)$ given by \eqref{coef-B}.
\subsubsection{Null tetrad adapted to the geometry of the matter distribution and curvature components}
In \cite{Gallo-lens-2011} Gallo and Moreschi constructed a null tetrad in terms of the metric components $A(r)$ and $B(r)$ for a static, spherically symmetric and asymptotically flat spacetime with a metric like \eqref{metric-spherically-symm}.
In order to include the effects of the plasma we can repeat those calculations to obtain a principal null tetrad $\{\tilde{\ell}_P^a,\tilde{n}_P^a,\tilde{m}_P^a,\tilde{\bar{m}}_P^a\}$ associated to the Gordon-like metric \eqref{effective-metric-2} in terms of the coefficients $\tilde{A}(r)$ and $\tilde{B}(r)$. Explicitly we have,
\begin{eqnarray}
\tilde{\ell}_P^a &=& \frac{1}{\tilde{A}}\bigg(\frac{\partial}{\partial\tilde{t}}\bigg)^a+\frac{1}{\sqrt{\tilde{A}\tilde{B}}}\bigg(\frac{\partial}{\partial r}\bigg)^a, \\
\tilde{n}_P^a &=& \frac{1}{2}\bigg[ \bigg(\frac{\partial}{\partial\tilde{t}}\bigg)^a- \sqrt{\frac{\tilde{A}}{\tilde{B}}}\bigg(\frac{\partial}{\partial r}\bigg)^a\bigg], \\
\tilde{m}_P^a &=&\frac{e^{i\varphi}}{\sqrt{2}r} \bigg[ \bigg(\frac{\partial}{\partial\theta}\bigg)^a - \frac{i}{\sin(\theta)} \bigg(\frac{\partial}{\partial\varphi}\bigg)^a\bigg],
\end{eqnarray}
where the coefficients $\tilde{A}(r)$ and $\tilde{B}(r)$ are given by \eqref{coef-A} and \eqref{coef-B}, respectively. However, let us note that in order to compute the leading linear order contributions to  the associated curvature scalars we only need to consider the previous principal null tetrad  at zero order, that is, we can set $\tilde{A}=\tilde{B}=1$.

As it was shown in \cite{Gallo-lens-2011}, at linear order, the transformation between the null tetrad $\{\ell^a,m^a,\bar{m}^a,n^{a}\}$ adapted to the motion of the light rays under study and the principal null tetrad adapted to the energy-momentum distribution induces the following transformation on the curvature scalars $\Phi_{00}$ and $\Psi_0$,
\begin{equation}\label{Psi0-transf}
\begin{split}
\Psi_0= 3\frac{b^2}{r^2}\tilde{\Psi}_2(r)e^{2i\vartheta},
\end{split}
\end{equation}
\begin{equation}\label{Phi00-transf}
\begin{split}
\Phi_{00}= \frac{2 b^2}{r^2}(\tilde\Phi_{11}-\frac{1}{4}\tilde\Phi_{00}) + \tilde\Phi_{00},
\end{split}
\end{equation}
where $\tilde{\Psi}_2$, $\tilde{\Phi}_{00}$ and $\tilde{\Phi}_{11}$ are the curvature scalars calculated from the Gordon-like metric \eqref{effective-metric-2} and using the tetrad $\{\tilde{\ell}_P^a,\tilde{n}_P^a,\tilde{m}_P^a,\tilde{\bar{m}}_P^a\}$, at first order. Explicitly, they are given by,
\begin{equation}
    \tilde{\Phi}_{00}=\frac{rM^{\prime}(r)-M(r)}{r^3}+\frac{1}{r\tilde{n}_{o}^{2}}\bigg(\frac{K_{e}N_{1}^{\prime}(r)}{2\omega_{\infty}^2}+\Phi^{\prime}(r)\bigg),
\end{equation}
\begin{equation}
    \tilde{\Phi}_{11}=\frac{M(r)}{2r^3}+\frac{1}{4\tilde{n}_{o}^2}\bigg( \frac{K_e N_{1}^{\prime\prime}(r)}{2\omega_{\infty}^2}+\Phi^{\prime\prime}(r) \bigg),
\end{equation}
\begin{equation}
\begin{aligned}
\tilde{\Psi}_2 =&\frac{rM^{\prime}(r)-3M(r)}{6r^3}
   -\frac{1}{6r\tilde{n}_{o}^2}\bigg( \frac{K_e}{2\omega_{\infty}^2} (N^{\prime}_{1}(r)\\
   &-rN^{\prime\prime}_{1}(r))+\Phi^{\prime}(r)-r\Phi^{\prime\prime}(r) \bigg).
\end{aligned}    
\end{equation}

Using the Einstein field equations \eqref{eq:rho}, \eqref{eq:mgrt} and the conservation equation \eqref{eq:consrt} it is possible to express these curvature components in terms of the energy-momentum tensor components and the function $M(r)$, which because of \eqref{eq:rho} contains the same information that $\varrho(r)$,
\begin{equation}
\begin{aligned}
    \tilde{\Phi}_{00}=&4\pi\bigg(\varrho(r)+\frac{P_{r}(r)}{\tilde{n}_{o}^2}\bigg)+\frac{M(r)}{r^3}\bigg( 1-\frac{1}{\tilde{n}_{o}^2} \bigg)\\ &+\frac{K_e}{2r\omega_{\infty}^2 \tilde{n}_{o}^2}N^{\prime}_{1}(r),
\end{aligned}    
\end{equation}

\begin{equation}
\begin{aligned}
    \tilde{\Phi}_{11}=&\frac{\pi}{\tilde{n}_{o}^2}\bigg( \varrho(r)+2P_{t}(r)-P_{r}(r) \bigg)+\frac{M(r)}{2r^3}\bigg(1-\frac{1}{\tilde{n}_{o}^2}\bigg)\\
    &+\frac{K_e}{8\omega_{\infty}^2 \tilde{n}_{o}^2}N^{\prime\prime}_{1}(r),
\end{aligned}    
\end{equation}

\begin{equation}
\begin{aligned}
\tilde{\Psi}_{2}=& \frac{4\pi}{3\tilde{n}_{o}^2}\bigg(P_{t}(r)-P_{r}(r)\bigg)+\bigg( \frac{2}{3}\pi\varrho(r)-\frac{M(r)}{2r^3} \bigg)\\
&\times\bigg( 1+\frac{1}{\tilde{n}_{o}^2} \bigg)
+ \frac{K_e}{12r\omega_{\infty}^2 \tilde{n}_{o}^2}\bigg( rN^{\prime\prime}_{1}(r)-N^{\prime}_{1}(r) \bigg).
\end{aligned}    
\end{equation}

Relations \eqref{Psi0-transf} and \eqref{Phi00-transf} together with the previous expressions for the curvature scalars allow us to write the optical scalars and the deflection angle in terms of the energy-momentum tensor components and the electronic density profile. In particular, from the relations \eqref{convergence}, \eqref{shear} and \eqref{alpha-kappa-gamma} we finally obtain,
\begin{equation}\label{convergence-Gallo1}
\begin{aligned}
\tilde{\kappa}=&\int_{-\lambda_{l}}^{\lambda_{ls}}\bigg[4\pi\bigg(\varrho+\frac{P_r}{\tilde{n}_{o}^2}\bigg)+\frac{4\pi b^2}{\tilde{n}_{o}^2 r^2}(P_t-P_r)\\
&+ \frac{1}{r^2}\bigg(\frac{3b^2}{2r^3}M-\frac{M}{r}-2\pi b^2\varrho\bigg)\bigg(1-\frac{1}{\tilde{n}_{o}^2}\bigg)\\
&-\frac{b^2 K_e}{4r^3 \omega_{\infty}^2 \tilde{n}_{o}^2}( N^{\prime}_{1} -rN^{\prime\prime}_{1})+\frac{K_e}{2r\omega_{\infty}^2 \tilde{n}_{o}^2}N^{\prime}_{1}
\bigg]dy,
\end{aligned}    
\end{equation}

\begin{equation}\label{shear-Gallo1}
\begin{aligned}
\tilde{\gamma}=& \int_{-\lambda_{l}}^{\lambda_{ls}}\frac{b^2}{r^2}\bigg[\frac{4\pi}{\tilde{n}_{o}^2}(P_{r}-P_{t})+\bigg( \frac{3}{2}\frac{M}{r^3}-2\pi\varrho \bigg)\\
&\times\bigg( 1+\frac{1}{\tilde{n}_{o}^2} \bigg)
+ \frac{K_e}{4r\omega_{\infty}^2 \tilde{n}_{o}^2}( N^{\prime}_{1} -rN^{\prime\prime}_{1})\bigg]dy,
\end{aligned}    
\end{equation}

\begin{equation}\label{angle-Gallo1}
\begin{aligned}
\alpha=& b\int_{-\lambda_{l}}^{\lambda_{ls}}\bigg[ 4\pi\bigg( \varrho+\frac{P_r}{\tilde{n}_{o}^2} \bigg)+\frac{b^2}{r^2}\bigg( \frac{3M}{r^3}-4\pi\varrho \bigg)\\
&-\frac{M}{r^3}\bigg(1-\frac{1}{\tilde{n}_{o}^2}\bigg)+\frac{K_e N^{\prime}_{1}}{2r\omega_{\infty}^2 \tilde{n}_{o}^2}\bigg]dy,
\end{aligned}    
\end{equation}
where we are integrating in the $y$ coordinate defined in Eq.\eqref{eq:parametrizaci}.

As before, $\lambda_l$ denotes the distance from the observer to the center of the lens if the spacetime were flat. In a similar way, $\lambda_{ls}=\lambda_s-\lambda_l$. These expressions generalize the results obtained in \cite{Gallo-lens-2011} for pure gravity ($\tilde{n}_{o}=1$ and $N_1=0$). In particular it is easy to see that in that limit Eq.\eqref{angle-Gallo1} reduces to \eqref{angle-Gallointro}.

\subsubsection{$\lambda_{ls}\to \infty$ and $\lambda_{l}\to \infty$}

In typical astrophysical situations, it is often convenient to assume that both the observer and the source are far away from the lens. In this case, by replacing the extremes of integration $\lambda_{ls}\to \infty$ and $\lambda_{l}\to \infty$ and integrating by parts,  we can show that the explicit dependence on $\varrho(r)$ can be omitted in \eqref{convergence-Gallo1}, \eqref{shear-Gallo1} and \eqref{angle-Gallo1}.

By using the Einstein field equation \eqref{eq:rho} we can write $\varrho(r)$ as a first derivative of $M(r)$ and then integrate by parts all the terms containing $\varrho(r)$. Doing that, it is easy to check the identities,
\begin{equation}\label{rho-1}
\int_{-\infty}^{\infty}\varrho(r)dy =\frac{1}{4\pi}\int_{-\infty}^{\infty}\frac{M(r)}{r^3}\bigg(1-\frac{r^2}{b^2-r^2}\bigg)dy,
\end{equation}
\begin{equation}\label{rho-2}
\int_{-\infty}^{\infty}\frac{\varrho(r)}{r^2}dy =\frac{1}{4\pi}\int_{-\infty}^{\infty}\frac{M(r)}{r^5}\bigg(3-\frac{r^2}{b^2-r^2}\bigg)dy,
\end{equation}
where the boundary terms vanish due to the spacetime is asymptotically flat which implies,
\begin{equation}
    \lim_{y\to\pm\infty} \frac{M(r)}{r^2}=0 \ \ \text{and} \ \ \lim_{y\to\pm\infty} \frac{M(r)}{r^4}=0.
\end{equation}

Using the relations \eqref{rho-1} and \eqref{rho-2} one can express the optical scalars and the deflection angle as,
\begin{equation}
\begin{aligned}
\tilde{\kappa}=&\int_{-\infty}^{\infty}\bigg[\frac{4\pi}{\tilde{n}_{o}^2}\bigg(P_r+\frac{b^2}{r^2}(P_t-P_r)\bigg)\\
&+ \frac{M}{2r^3}\bigg(1-\frac{r^2}{b^2-r^2}\bigg)\bigg(1+\frac{1}{\tilde{n}_{o}^2}\bigg)\\
&-\frac{b^2 K_e}{4r^3 \omega_{\infty}^2 \tilde{n}_{o}^2}( N^{\prime}_{1} -rN^{\prime\prime}_{1})+\frac{K_e}{2r\omega_{\infty}^2 \tilde{n}_{o}^2}N^{\prime}_{1}
\bigg]dy,
\end{aligned}    
\end{equation}

\begin{equation}
\begin{aligned}
\tilde{\gamma}=& \int_{-\infty}^{\infty}\frac{b^2}{r^2}\bigg[\frac{4\pi}{\tilde{n}_{o}^2}(P_{r}-P_{t})+\frac{M}{2r}\bigg(\frac{1}{b^2-r^2}\bigg)\\
&\times\bigg( 1+\frac{1}{\tilde{n}_{o}^2} \bigg)
+ \frac{K_e}{4r\omega_{\infty}^2 \tilde{n}_{o}^2}( N^{\prime}_{1} -rN^{\prime\prime}_{1})\bigg]dy,
\end{aligned}    
\end{equation}

\begin{equation}\label{angle-Gallo2}
    \alpha=\int_{-\infty}^{\infty} \frac{b}{r}\bigg[  \frac{M}{r^2}\bigg(1+\frac{1}{\tilde{n}_{o}^{2}}\bigg)+\frac{4\pi r P_{r}}{\tilde{n}^2_o}+\frac{K_e N^{\prime}_1}{2\omega_{\infty}^2 \tilde{n}_{o}^2}\bigg] dy.
\end{equation}
It is worthwhile to notice that even when these expressions are very compact and they are potentially very practical, we have no knowledge about a previous presentation of them in the literature.

\subsubsection{Analogy between photons in an homogeneous plasma and massive particles in pure gravity}

As is very well known there exists a correspondence between the motion of photons in an homogeneous plasma ($N^{\prime}_1=0$) in a given spacetime and the geodesic motion of massive particles on the same background.

In the context of this work, this correspondence reduces to setting $N^{\prime}_1=0$ and to identify $\tilde{n}_o$ with the speed $v$ (which represents the initial speed of a test massive particle as measured by an asymptotic static observer when the particle was far away from the gravitational source). For more details see \cite{Kulsrud-1992,BisnovatyiKogan:2010ar,Crisnejo-gauss-bonnet-1}. Then, the deflection angle $\alpha_{\text{mp}}$ for massive particles is given by
\begin{equation}\label{angle-Gallo22}
    \alpha_{\text{mp}}=\int_{-\infty}^{\infty} \frac{b}{r}\bigg[  \frac{M}{r^2}\bigg(1+\frac{1}{v^{2}}\bigg)+\frac{4\pi r P_{r}}{v^2}\bigg] dy.
\end{equation}
Alternatively it can be written as,
\begin{equation}\label{alpha-particles}
\alpha_{\text{mp}}=\frac{\alpha_{M}}{2}\bigg(1+\frac{1}{v^{2}}\bigg)+\frac{\alpha_{P_r}}{v^2},
\end{equation}
with
\begin{equation}
    \alpha_M = 2\int_{-\infty}^\infty \frac{bM}{r^3} dy, \ \ \ \alpha_{P_r}=4\pi\int_{-\infty}^{\infty}r P_r dy.
\end{equation}
The expression \eqref{alpha-particles} generalizes the results found in \cite{Gallo-lens-2011} by including the deflection of massive particles. In particular, if $v=1$, that is, if we consider massless particles, then the deflection angle reduces to $\alpha=\alpha_M+\alpha_{P_r}$, and therefore these quantities represent the contribution to the deflection angle for massless particles due to the total mass and the radial pressure. Note that these results are independent of the particular mass density profile under consideration. In particular, if the spacetime is such that the pressure can be neglected, the relation between the deflection angle for massive $\alpha_\text{mp}$ and massless particles $\alpha_\gamma\equiv\alpha_M$ is given by 
\begin{equation}
\alpha_{\text{mp}}=\frac{\alpha_\gamma}{2}\bigg(1+\frac{1}{v^2}\bigg).
\end{equation}

\subsection{Using the Gauss-Bonnet theorem}

In order to check the results found in the previous subsection we will use the geometrical method developed by Gibbons and Werner \cite{Gibbons-gauss-bonnet} and extended by us to include the plasma effects \cite{Crisnejo-gauss-bonnet-1}. This method of studying gravitational lenses is based on the use of the Gauss-Bonnet theorem.

As we have seen in \cite{Crisnejo-gauss-bonnet-1} for a spacetime $(\mathcal{M},g_{\alpha\beta})$ with line element given by \eqref{metric-spherically-symm}, spatial projections of the light propagating through a medium with refractive index $n$ on the slices $t=\text{constant}$ are geodesics of the following Riemannian optical metric,
\begin{equation}
    d\sigma^2=g^{\text{opt}}_{ij} dx^i dx^j=\frac{n^2(r)}{A(r)}\bigg(B(r)dr^2+r^2 d\varphi^2 \bigg),
\end{equation}
where we have restricted to the equatorial plane $\vartheta=\pi/2$ without loss of generality due to the spherical symmetry. Again, we express the metric functions $A(r)$ and $B(r)$ as,
\begin{equation}
    A(r)=e^{2\Phi(r)}, \ \ \ B(r)=\frac{1}{1-\frac{2M(r)}{r}}.
\end{equation}
For a cold nonmagnetized plasma the refractive index given by \eqref{eq:indrefp}
can be equivalently rewritten as
\begin{equation}
    n^2(r)=1-(1-n_{o}^{2}(r))A(r),
\end{equation}
where the redshift has already taken into account and $n_{o}^{2}(r)$ is given by,
\begin{equation}
    n_{o}^{2}(r)=1-\frac{\omega_{e}^2(r)}{\omega^{2}_\infty}.
\end{equation}

\begin{figure}[h]
\centering
\includegraphics[clip,width=75mm]{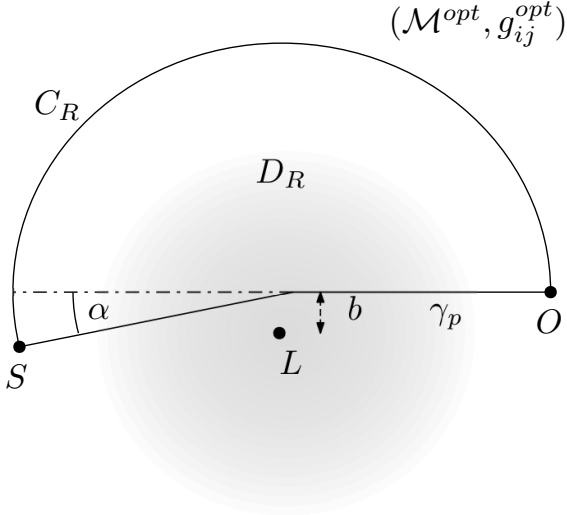}
\caption{The points $S$ and $O$ represent the source and observer position while $\gamma_p$ indicates a light ray emitted by the source and that reaches the observer. b is identified with the impact parameter.}
 \label{plot-gb}
\end{figure}

In order to calculate the deflection angle using the Gauss-Bonnet theorem we have to choose a specific domain in the optical manifold $(\mathcal{M}^{\text{opt}},g_{ij}^{\text{opt}})$. There are several ways to do that, the most simple domain, and following the original work of Gibbons and Werner, is to consider a simply connected region as shown in Fig. \ref{plot-gb}, where its boundary is formed by the spatial geodesic $\gamma_p$ followed by the photon in the optical manifold and the curve $C_R$ defined as $r(\varphi)=R=\text{constant}$. Then, the deflection angle can be computed by  \cite{Gibbons-gauss-bonnet,Crisnejo-gauss-bonnet-1},

\begin{equation}\label{Gauss-Bonnet-equation}
    \lim_{R\to\infty}\int_0^{\pi+\alpha}\bigg[\kappa_g \frac{d\sigma}{d\varphi}\bigg]\bigg|_{C_R}d\varphi=\pi-\lim_{R\to\infty}\int\int_{D_R}\mathcal{K}dS,
\end{equation}
where $\mathcal{K}$ is the Gaussian curvature, $dS$ is the surface element in coordinates $(r, \vartheta)$ and $\kappa_g$ is the geodesic curvature of $C_R$.

As we are interested only on the linear corrections in $M(r)$ and $\Phi(r)$ to the deflection angle, it is sufficient to consider $r_{\varphi}$ as the background solution,
\begin{equation}
    r_{\varphi}=\frac{b}{\sin(\varphi)},
\end{equation}
and, on the other hand, we only need to compute $\mathcal{K}dS$ and $\kappa_g \frac{d\sigma}{d\varphi}$ at first order in $M(r)$ and $\Phi(r)$,
\begin{equation}\label{eq:kdsecn}
\begin{aligned}
    \mathcal{K}dS=&\bigg[\frac{(r \Phi^{\prime})^{\prime}}{n_{o}^2}+\bigg(\frac{M}{r}\bigg)^{\prime} -\frac{(r n_{o}^{\prime})^{\prime}}{n_{o}}+ \mathcal{F}\bigg] dr d\varphi,
\end{aligned}    
\end{equation}
where
\begin{equation}
\begin{aligned}
    \mathcal{F}=& \mathcal{F}(\Phi n_{o}^{\prime}, \Phi n_{o}^{\prime\prime},\Phi^{\prime} n_{o}^{\prime},(n_{o}^{\prime})^{2},M n_{o}^{\prime},M n_{o}^{\prime\prime},M^{\prime}n_{o}^{\prime})\\
    =& r \frac{(n_o^{\prime})^2}{n_o^4}-2\Phi \frac{(r n_{o}^{\prime})^{\prime}}{n_o^3}+6\Phi r \frac{(n_{o}^{\prime})^2}{n_o^4}
    -4\Phi^{\prime}r \frac{n_{o}^{\prime}}{n_o^3} \\
    &+  \frac{(n_{o}^{\prime}M)^{\prime}}{n_o}- \frac{(n_{o}^{\prime})^{2} M}{n_o^2},
\end{aligned}    
\end{equation}
and $\,^\prime$ denotes derivative respect to the radial coordinate. On the other hand,
\begin{equation}
    \kappa_g \frac{d\sigma}{d\varphi}=1-\frac{r\Phi^{\prime}}{n_o^2}-\frac{M}{r}+\frac{r n^{\prime}_o}{n_o} + \mathcal{G},
\end{equation}
where $\mathcal{G}=\mathcal{G}(\Phi n^{\prime}_o,M n^{\prime}_o)$ is given by
\begin{equation}
    \mathcal{G}= 2\frac{r\Phi  n^{\prime}_o}{n^3_o}-M\frac{n^{\prime}_o}{n_o}.
\end{equation}

In the following, we will neglect the contributions plasma$\times$gravity given by the terms $\mathcal{F}$ and $\mathcal{G}$. Now let us consider two situations: one in which the two-form $\mathcal{K}dS\ne0$ and the other when it vanishes. 

\subsubsection{$\mathcal{K}dS\ne0$}
If $\mathcal{K}dS\ne0$ we get the following expression for the deflection angle,
\begin{equation}\label{angle-bonnet-surface}
    \alpha=-\int_{0}^{\pi}\int_{r_{\varphi}}^{\infty}\bigg[\frac{(r \Phi^{\prime})^{\prime}}{n_{o}^2}+\bigg(\frac{M}{r}\bigg)^{\prime}-\frac{(r n_{o}^{\prime})^{\prime}}{n_{o}} \bigg] dr d\varphi.
\end{equation}
By integrating by parts and discarding terms of order $\mathcal{O}((n_{o}^{\prime})^2,\Phi^{\prime}n_{o}^{\prime})$, we can rewrite \eqref{angle-bonnet-surface} as a line integral
\begin{equation}
    \alpha = \int_{0}^{\pi}\bigg[\frac{r \Phi^{\prime}}{n_{o}^2}+\frac{M}{r}-\frac{r n_{o}^{\prime}}{n_{o}}\bigg]\bigg|_{r=r_{\varphi}}d\varphi,
\end{equation}
where we have assumed that
\begin{equation}
    \lim_{r\to\infty} \frac{r n_{o}^{\prime}}{n_{o}}=0 \ \ \ \ \ \text{and} \ \ \ \ \ \lim_{r\to\infty} \frac{r \Phi^{\prime}}{n_{o}^2}=0,
\end{equation}

We transform to a new coordinate $y$ related to $r$ by $y=\sqrt{r^2-b^2}$, thus satisfying $\tan\varphi=b/y$. By using the Einstein equation at first order in $M$:
\begin{equation}
    \Phi^{\prime}=\frac{M(r)}{r^2}+4\pi r P_{r}(r),    
\end{equation}
we finally obtain an expression for the deflection angle in terms of components of the energy-momentum tensor for a lens surrounded by plasma,
\begin{equation}\label{eq:angulon0}
    \alpha=\int_{-\infty}^{\infty} \frac{b}{r}\bigg[ \frac{M}{r^2}\bigg(1+\frac{1}{n_{o}^{2}}\bigg)+\frac{4\pi r P_{r}}{n^2_o}-\frac{n_{o}^{\prime}}{n_{o}}\bigg]\bigg|_{r=r_{y}} dy
\end{equation}
where $r_{y}=\sqrt{b^2+y^2}$.

\subsubsection{$\mathcal{K}dS=0$}
This situation could be found for a specific kind of plasma media. For the case that we are in a pure gravity situation, this condition is found for example when the mass distribution is determined by an isothermal density profile\cite{Gibbons-gauss-bonnet}.
If $\mathcal{K}dS=0$ then by discarding the $\mathcal{F}$ term and also all the terms of order $\mathcal{O}((n_{o}^{\prime})^2,\Phi^{\prime}n_{o}^{\prime})$, we can see from Eq.\eqref{eq:kdsecn} that
\begin{equation}
\begin{split}
    &\bigg(\frac{r \Phi^{\prime}}{n_{o}^2}+\frac{M}{r}-\frac{r n_{o}^{\prime}}{n_{o}}\bigg)' = \mathcal{O}((n_{o}^{\prime})^2,\Phi^{\prime}n_{o}^{\prime})-\mathcal{F}\\
    &=\mathcal{O}\bigg(\Phi n_{o}^{\prime}, \Phi n_{o}^{\prime\prime},\Phi^{\prime} n_{o}^{\prime},(n_{o}^{\prime})^{2},M n_{o}^{\prime},M n_{o}^{\prime\prime},M^{\prime}n_{o}^{\prime}\bigg);
    \end{split}
\end{equation}
which implies that the following quantity is constant with respect to the radial coordinate at the considered order
\begin{equation}
    \frac{r \Phi^{\prime}}{n_{o}^2}+\frac{M}{r}-\frac{r n_{o}^{\prime}}{n_{o}} = C=\text{constant}.
\end{equation}
It should be noted that this constant $C$ is a first order quantity.
Therefore it follows that at this order
\begin{equation}
\begin{aligned}
    \bigg[\kappa_g\frac{d\sigma}{d\varphi}\bigg]\bigg|_{C_R}
    &=\bigg[1-\frac{r\Phi^{\prime}}{n_o^2}-\frac{M}{r}+\frac{r n^{\prime}_o}{n_o}\bigg]\bigg|_{R(\varphi)} \\
    &=\bigg[1-\frac{r\Phi^{\prime}}{n_o^2}-\frac{M}{r}+\frac{r n^{\prime}_o}{n_o}\bigg]\bigg|_{r_\varphi}.
\end{aligned}
\end{equation}
where we have used the fact that as it is constant, it can be evaluated along the trajectory $r_\varphi=b/\sin(\varphi)$ instead of $R(\varphi)=\text{constant}$.
Then, the relation \eqref{Gauss-Bonnet-equation} for calculating the bending angle reduces to,
\begin{equation}
    \int_0^{\pi+\alpha} \bigg[1-\frac{r\Phi^{\prime}}{n_o^2}-\frac{M}{r}+\frac{r n^{\prime}_o}{n_o}\bigg]\bigg|_{r_\varphi} d\varphi=\pi.
\end{equation}
Finally, by splitting the domain of integration from $0$ to $\pi$ and from $\pi$ to $\pi+\alpha$ and approximating,
\begin{equation}
    \int_\pi^{\pi+\alpha} \bigg[1-\frac{r\Phi^{\prime}}{n_o^2}-\frac{M}{r}+\frac{r n^{\prime}_o}{n_o}\bigg]\bigg|_{r_\varphi} d\varphi\approx\alpha,
\end{equation}
we obtain
\begin{equation}
    \alpha = \int_{0}^{\pi}\bigg[\frac{r \Phi^{\prime}}{n_{o}^2}+\frac{M}{r}-\frac{r n_{o}^{\prime}}{n_{o}}\bigg]\bigg|_{r=r_{\varphi}}d\varphi=C\pi.
\end{equation}
Note that the integral expression coincides with the expression for $\alpha$ in the $\mathcal{K}dS \ne 0$ case. That is, for the practical purpose it is not necessary to discriminate between the two cases and we can use the same formula for the bending angle.

\subsubsection{Comparison}
Additionally, if we split $\omega_e^2$ as in Eq.\eqref{omega-plasma-comp-1} we get
\begin{equation}
n_{o}^{2}=\tilde{n}_{o}^2\bigg(1-\frac{K_e N_1(r)}{\tilde{n}_{o}^2\omega_{\infty}^2}\bigg)
\end{equation}
and then
\begin{equation}
    -\frac{n_{o}^{\prime}}{n_{o}}=\frac{K_{e} N_{1}^{\prime}(r)}{2n_{o}^{2}\omega_{\infty}^{2}}.
\end{equation}
On the other hand, because of we are assuming that $\frac{K_e N_{1}(r)}{\omega_{e0}^2}\ll 1$, we  can approximate $n_{o}^2\approx\tilde{n}_{o}^2$, obtaining finally,
\begin{equation}\label{eq:121}
    \alpha=\int_{-\infty}^{\infty} \frac{b}{r}\bigg[  \frac{M}{r^2}\bigg(1+\frac{1}{\tilde{n}_{o}^{2}}\bigg)+\frac{4\pi r P_{r}}{\tilde{n}^2_o}+\frac{K_e N^{\prime}_1}{2\omega_{\infty}^2 \tilde{n}_{o}^2}\bigg] dy,
\end{equation}
which completely coincides with \eqref{angle-Gallo2}.
In this way, we have obtained the same expression for the deflection angle in terms of the energy-momentum tensor components using two different geometrical approaches.
 
\section{Correspondence between the spatial motion of photons in a nonhomogeneous plasma and massive particles in a gravitational and external field}\label{analogy-plasma-charged-particle}

As was discussed above, there exists a well known correspondence between the geodesic motion of test massive particles in a gravitational field and the motion of photons in a homogeneous cold plasma medium.  
In this section we show that there exist also a striking  correspondence between the nongeodesic motion of test charged massive particles in a gravitational field where an electrical field is present and the motion of photons in a non homogeneous cold nonmagnetized plasma.

Let us consider a static spacetime whose line element is given by,
\begin{equation}
    ds^2=A(x^i)dt^2-g_{ij}dx^i dx^j, \ \ i,j,k,... = 1, 2, 3.
\end{equation}
The action for a test charged massive particle with charge $q$ and mass $\mu$ moving under the influence of the gravitational field and a static electrical field determined by a potential $U(x^i)$ is given by
\begin{equation}
    \mathcal{S}=\int \mathcal{L}(x^i,\dot{x}^i) dt,
\end{equation}
where the integral is along the world line of the particle and the Lagrangian density $\mathcal{L}$ reads,
\begin{equation}
    \mathcal{L}(x^i,\dot{x}^i)=-\mu\sqrt{A(x^i)-g_{ij}\dot{x}^i\dot{x}^j}-qU(x^i).
\end{equation}

Before continuing, we recall that the most general Lagrangian for a test massive charged particle in an Einstein-Maxwell field is given by $\mathcal{L}=-\mu\sqrt{g_{\alpha\beta}u^\alpha u^\beta}-qA_\alpha u^\alpha$.  Here we are assuming that in the $\{t,x^i\}$ coordinates only $A_0\neq 0$. More importantly, our considerations will be also valid for \emph{any} central scalar potential $U(x^i)$ not necessarily of electromagnetic nature.

The motion of the particle can be studied from the corresponding Euler-Lagrange equations or equivalently from the Hamilton-Jacobi equations of the Hamiltonian $\mathcal{H}(x^i,p_i)$ given by the Legendre transformation,
\begin{equation}
    \mathcal{H}(x^i,p_i)=p_i \dot{x}^i-\mathcal{L}(x^i,\dot{x}^j),
\end{equation}
where $\dot{x}^i=\dot{x}^i(x^j,p_k)$. The 4-momentum is defined as,
\begin{equation}
    p_i=\frac{\partial \mathcal{L}}{\partial \dot{x}^i}.
\end{equation}
For this case the Hamiltonian takes the form,
\begin{equation}\label{Hamilt-charged-massive}
    \mathcal{H}(x^i,p_i)=\sqrt{\mu^2A(x^i)+A(x^i)g^{ij}p_i p_j}+qU(x^i).
\end{equation}

A few years ago, Gibbons introduced the Jacobi metric formulation to study the motion of test massive particles in static spacetimes \cite{Gibbons-Jacobi-static}. In this formulation the motion of these particles is given by the geodesics of a energy-dependent Riemmanian metric known as the Jacobi metric. 

Moreover, any nongeodesic motion can be described by a geodesic associated with the Jacobi metric derived from a given Hamiltonian. In particular, for a test particle whose motion is described by the Hamiltonian \eqref{Hamilt-charged-massive} the Jacobi metric $J_{ij}$ is given by,
\begin{equation}
    J_{ij}=E_{\infty}^2 \; \hat{g}_{ij}^{\text{opt}},
\end{equation}
where
\begin{equation}\label{analog-prefactor}
    \hat{g}_{ij}^{\text{opt}}= \bigg[ \bigg( 1-\frac{qU(x^i)}{E_{\infty}} \bigg)^2 -\frac{\mu^2 A(x^i)}{E_{\infty}^2} \bigg] \frac{g_{ij}}{A(x^i)}.
\end{equation}
For a free massless test particle, $\hat{g}_{ij}^{\text{opt}}$ coincides with the optical metric $g_{ij}^{\text{opt}}$ in pure gravity.

On the other hand, we know that the photon trajectory in a medium described by a refractive index $n$ are geodesics with respect to the optical metric
\begin{equation}\label{eq:optmcp}
g_{ij}^{\text{opt}}= n^2(x^i)\frac{g_{ij}}{A(x^i)},
\end{equation}
which in particular for a cold nonmagnetized plasma $n$ is given by \eqref{eq:indrefp}.

By splitting the electronic density profile $\omega_e^2$ as,
\begin{equation}\label{electronic-profile-charged}
    \omega_{e}^2(x^i)=\omega_{eo}^2+K_e N_1(x^i),
\end{equation}
with $\omega_{eo}=\text{constant}$ and comparing the refractive index $n$ with the factor that precedes to $\frac{g_{ij}}{A(x^i)}$ in \eqref{analog-prefactor} we can see that if we make the following identifications
\begin{gather}
\omega_{eo} \leftrightarrow \mu,  \ \ \omega_{\infty} \leftrightarrow E_{\infty} , \label{identification-1} \\
N_1(x^i) \leftrightarrow \frac{qU(x^i)}{K_e A(x^i)}(2E_{\infty}-qU(x^i)), \label{identification-2}
\end{gather}
and $N_1(x^i)> 0$,
then 
the spatial orbit of a charged massive particle in a given spacetime when an electrical field is present is equivalent to the spatial orbit of a photon in a very particular (energy dependent) nonhomogeneous plasma, and vice-versa. Note that this analogy is not only valid when a gravitational field is present, it also occurs in a Minkowski spacetime. For example, if we assume that we have a fixed electric charge $Q$ generating a central potential $U(r)=Q/r$, the orbit of a test particle of mass $\mu$, charge $q$ and total energy $E_\infty$ governed by the Lorentz force agrees with the orbit of a photon with the same energy moving in a plasma which in addition to the homogeneous part of the plasma electronic number density given by $K_e\mu^2$ has also a non-uniform charge density that must be given by $\frac{qQ}{K_e r}(2E_{\infty}-\frac{qQ}{r})$, which will be positive if $qQ>0$, that is the repulsive character between charges can be described by the divergent effect of the trajectory of a particular photon with energy $E_\infty$ in a particular plasma density profile. Of course, for the rest of the photons with different energies moving in the same plasma medium, the correspondence is not achieved.

On the other hand, the Jacobi metric not only allows us to make the previous analogy between the motion of a photon in a cold plasma medium and particles in external fields, it also allows us to use the powerful Gauss-Bonnet theorem to study the scattering of relativistic particles 
 following nongeodesic motion, extending in this way the original purpose of this method which was only restricted to study of null geodesics. More precisely, for a spherically symmetric spacetime we can use the expression \eqref{eq:angulon0} with $n_o(r)$ given by
 \begin{equation}
     n^2_o(r)=1-\frac{1}{E^2_\infty}\bigg(\mu^2+\frac{2qU(r)E_\infty}{A(r)}-\frac{q^2U(r)^2}{A(r)}\bigg),
 \end{equation}
 in order to compute the deflection angle. Keeping only the leading order terms of that expression, we obtain that the deflection angle reduces to \eqref{eq:121} which if we only keep the linear terms in $U(r)$ gives:
 \begin{equation}\label{eq:alphalinno}
 \begin{aligned}
 \alpha
 \approx & \int_{-\infty}^{\infty} \frac{b}{r}\bigg[ \frac{M}{r^2}\bigg(1+\frac{1}{v^{2}}\bigg)+\frac{4\pi r P_{r}}{v^2}+\frac{qU^{\prime}}{E_\infty v^2}\bigg]\bigg|_{r=r_{y}} dy.
  \end{aligned}
\end{equation}
This equation generalizes the expression found in \cite{Gallo-lens-2011} to a more general situation where massive particles and nongravitational external central potential are allowed.

In the following subsection we show the power of this technique by computing the deflection angle of relativistic charges particles in a Reissner-Nordst\"om spacetime keeping terms of higher order than in \eqref{eq:alphalinno}. For a detailed study of the Jacobi metric for this particular situation where a more general kind of orbit are discussed we refer to the recent work of Das, Sk and Ghosh \cite{Das:2016opi}.

\subsection{Deflection angle of charged massive particles in a Reissner-Nordstr\"{o}m spacetime}
Let us consider a test charged massive particle moving in a Reissner-Nordstr\"{o}m spacetime
\begin{equation}
\begin{aligned}
    ds^2=&\bigg(1-\frac{2m}{r}+\frac{Q^2}{r^2}\bigg)dt^2-\frac{dr^2}{1-\frac{2m}{r}+\frac{Q^2}{r^2}} \\
    &- r^2 (d\vartheta^2+\sin^2\vartheta d\varphi^2),
\end{aligned}    
\end{equation}
and under the action of the electric Coulomb potential
\begin{equation}\label{Coulomb-potential}
    U(r)=\frac{Q}{r}.
\end{equation}
The energy $E_\infty$ and angular momentum $J$ of the particle measured by a static asymptotic observer are given by the following expressions \cite{Crisnejo-gauss-bonnet-1},
\begin{equation}\label{Energy-angular-momentum}
E_\infty=\frac{\mu}{\sqrt{1-v^2}} \ \ \ \text{and} \ \ \ J=\frac{\mu v b}{\sqrt{1-v^2}}.
\end{equation}
In the above expression $v$ is the speed of the particle when it is in the asymptotic region as measured by the same observer while $b$ is the impact parameter.

Let us start by computing the linear order contributions to the deflection angle in terms of the energy-momentum tensor components and the potential as given by \eqref{eq:alphalinno}.  As is well known for a Reissner-Nordstr\"om metric the mass function $M(r)$ and the radial pressure $P_r(r)$ are given by
\begin{eqnarray}
    M(r)&=&m-\frac{Q^2}{2r},\\
    P_r(r)&=&-\frac{Q^2}{8\pi r^4}.
\end{eqnarray}
By replacing these relations and Eq.\eqref{Coulomb-potential} into \eqref{eq:alphalinno} and taking into account that $r=\sqrt{b^2+y^2}$ we find that at linear order the deflection angle is given by
\begin{equation}
    \alpha=\frac{2m}{b}\bigg(1+\frac{1}{v^2}\bigg)-\frac{\pi Q^2}{4b^2}\bigg( 1+\frac{2}{v^2}\bigg)-\frac{2qQ}{bv^2E_\infty}. 
\end{equation}
Note that we have obtained this expression in a very easy way through the use of the Eq.\eqref{eq:alphalinno}, which was derived in a geometrical way using the null-tetrad approach and the Gibbons-Werner method.

Let us move now to the computation of higher order corrections to the deflection angle. It has been proved in many works that the Gauss-Bonnet method is a very useful tool to calculate the deflection angle. In order to apply this method we can proceed in different ways. For instance, we can calculate the deflection angle of a photon moving in a nonhomogeneous plasma with electronic density profile given by \eqref{electronic-profile-charged} and then use the identifications \eqref{identification-1} and \eqref{identification-2} with the nongravitational potential $U$ given by \eqref{Coulomb-potential}. In that case the correspondence is only physical if the test particle has a charge with the same sign as the total charge of the black hole. However, the use of the optical metric as given by \eqref{eq:optmcp} is independent of the sign of the charges.
In our case the effective refractive index given by,
\begin{equation}\label{eq:nparausar}
    n^2(r)=1-\bigg(1-\frac{2m}{r}+\frac{Q^2}{r^2}\bigg)(1-v^2)-\frac{2qQ}{rE_\infty}+\frac{q^2Q^2}{r^2E_\infty^2},
\end{equation}
where we have used the first identity in \eqref{Energy-angular-momentum} in  order to express the refractive index in terms of the velocity, charge and energy of the particle. 

Now we can apply the Gauss-Bonnet method as developed in \cite{Crisnejo-gauss-bonnet-1} for the nonpure gravity case to compute the deflection angle. The optical metric reads,
\begin{equation}
    d\sigma^2=n^2(r)\bigg(\frac{dr^2}{(1-\frac{2m}{r}+\frac{Q^2}{r^2})^2}+\frac{r^2d\varphi^2}{1-\frac{2m}{r}+\frac{Q^2}{r^2}}\bigg).
\end{equation}
At this time it is convenient to introduce the following dimensionless parameter, which we assume are small, 
\begin{equation}\label{parameters-dimless}
\beta=\frac{m}{b}, \ \ \gamma=\frac{Q^2}{b^2}, \ \ \delta=\frac{qQ}{b^2}.
\end{equation}
We are interested in the computation of the deflection angle conserving all the terms of order  $\mathcal{O}(\beta,\gamma,\delta,\beta^2,\delta^2,\beta\delta)$; and therefore we need the following expression for the two-form $\mathcal{K}dS$,
\begin{equation}
\begin{aligned}
\mathcal{K}dS=&\bigg[-\frac{b\beta}{r^2}\bigg(1+\frac{1}{v^2}\bigg)+\frac{b^2\delta}{r^2v^2E_\infty}-(v^4+6v^2-4)\\
&\times \frac{b^2\beta^2}{r^3v^4}+2(3v^2-4)\frac{b^3\beta\delta}{r^3v^4E_\infty}+(2+v^2)\frac{b^2\gamma}{r^3 v^2}\\
&+(2-v^2)\frac{2b^4\delta^2}{r^3v^4 E^2_{\infty}}\bigg]d\varphi dr \\
&+ \mathcal{O}(\gamma^2,\beta^3,\delta^3,\delta\beta^2,\beta\gamma,\beta\delta^2,\delta\gamma),
\end{aligned}    
\end{equation}
or equivalently, in terms of the variable $u=1/r$ as
\begin{equation}
\begin{aligned}
\mathcal{K}dS=&\bigg[b\beta\bigg(1+\frac{1}{v^2}\bigg)-\frac{b^2\delta}{v^2E_\infty}+(v^4+6v^2-4)\\
&\times \frac{ub^2\beta^2}{v^4}-2(3v^2-4)\frac{ub^3\beta\delta}{v^4E_\infty}-(2+v^2)\frac{ub^2\gamma}{v^2}\\
&-(2-v^2)\frac{2ub^4\delta^2}{v^4 E^2_{\infty}}\bigg]d\varphi du \\
&+\mathcal{O}(\gamma^2,\beta^3,\delta^3,\delta\beta^2,\beta\gamma,\beta\delta^2,\delta\gamma).
\end{aligned}    
\end{equation}

In order to calculate the deflection angle to 
the desired order (which include terms of order $\beta^2$ and $\delta^2$) we have to integrate 
along a spatial orbit which takes into account 
corrections of order $\beta$ and $\delta$ to 
the straight line orbit (given by 
$u_0=\sin(\varphi)/b$). These corrections to 
the orbit are needed because even when  
$\mathcal{K}dS$ has already contributions of 
order $\beta^2$ and $\delta^2$,\footnote{These terms
will contribute to the deflection angle with 
$\beta^2$ and $\delta^2$ terms when the 
required integral is valued at the integration 
limit given by the zero order orbit $u_0$.} it also 
contains linear terms in these parameters, and 
therefore it follows that the integral of these linear terms valued in the $\beta$ and $\delta$ corrections to the orbit will also produce 
extra contributions to the required quadratic 
order in those parameters. As 
we show in Appendix \ref{app-RN} at the 
considered order the orbit is given by
    \begin{equation}
\begin{aligned}
u(\varphi)=&\frac{1}{b} \bigg[\sin\varphi + \bigg((1-\cos\varphi)+v^2(\cos^2\varphi-\cos\varphi)\bigg)\frac{\beta}{v^2}\\
&-(1-\cos\varphi)\sqrt{1-v^2}\frac{\delta}{v^2\mu}\bigg].
\end{aligned}
\end{equation}
It is important to remark again that in this case we only had to solve the orbit equation up to $\beta$ and $\delta$ orders. This is the main difference between the use of the Gauss-Bonnet method for calculating the deflection angle and the alternative way which follows from the explicit solution of the spatial orbit equation. In the last approach one must to solve the orbit equation to the same order as the one needed for the deflection angle, which is much more laborious and cumbersome. More details are given in Appendix \ref{app-RN}, where the evaluation of the deflection angle using the orbital equation approach is realized in detail.

Finally, using the relation \eqref{Gauss-Bonnet-equation} where the left-hand side reduces to $\alpha+\pi$ due to the spacetime is asymptotically flat we obtain the following expression for the deflection angle,
\begin{equation}\label{alpha_RN-1}
\begin{aligned}
    \alpha=&\frac{2m}{b}\bigg(1+\frac{1}{v^2}\bigg)+\frac{3\pi m^2}{4b^2}\bigg(1+\frac{4}{v^2}\bigg)-\frac{\pi Q^2}{4b^2}\bigg( 1+\frac{2}{v^2}\bigg)\\
    &-\frac{2qQ}{bv^2E_\infty}+\frac{\pi q^2 Q^2}{2b^2v^2E_\infty^2}-\frac{3\pi qQm}{b^2v^2E_\infty}. 
\end{aligned}    
\end{equation}
The first three terms agrees with the ones obtained recently by Pang and Jia \cite{Pang:2018jpm} where they studied the deflection angle of neutral massive particles in a Reissner-Nordst\"om background. The fourth and fifth terms are also present in a Minkowski spacetime. For that case, an exact expression for the deflection angle (valid for all kind of velocities, and magnitude of the deflection angle) was found by Synge (see Appendix C of \cite{book:synge}). In our notation his expression reads:
\begin{equation}\label{eq:alphasyn}
    \alpha=-\pi+\frac{4}{K}\arctan{\sqrt{\Gamma}};
\end{equation}
where 
\begin{equation}\label{eq:K}
    K=\sqrt{1-\frac{\delta^2b^2}{E^2_\infty-1}},
\end{equation}
and
\begin{equation}\label{eq:K2}
    \Gamma=\frac{\sqrt{\delta^2b^2\mu^2+(E^2_\infty-\mu^2)^2}-\delta E_\infty b}{\sqrt{\delta^2b^2\mu^2+(E^2_\infty-\mu^2)^2}+\delta E_\infty b},
\end{equation}
and $\delta$ is defined in \eqref{parameters-dimless}.

Taking into account Eqs.\eqref{eq:K} and \eqref{eq:K2}, and by doing a Taylor expansion of \eqref{eq:alphasyn} in terms of the parameter $\delta$ we obtain:
\begin{equation}
\alpha=-\frac{2E_\infty b}{E^2_\infty-\mu^2}\delta+\frac{b^2\pi}{2(E^2_\infty-\mu^2)}\delta^2+\mathcal{O}(\delta^3).    
\end{equation}
It is easy to check that this expression agrees with the forth and fifth terms of \eqref{alpha_RN-1}. The last term is a Lorentz force term which indicates that the dynamics are indeed in a curved background.

\section{Final Remarks}
In this paper we have shown how the formalism of null tetrads originally proposed to express the deflection angle for photons in pure gravity in terms of curvature scalars can be extended to be used when dispersive media are present. At the same time we have presented expressions for the optical scalars in terms of the different components of the momentum-energy tensor.

On the other hand we have arrived at several of the same results through the use of another recently proposed geometrical method that makes use of the Gauss-Bonnet theorem. In particular, although this method was originally proposed to study null geodesics, we have shown that it can be extended to the study of massive particles even when they do not follow timelike geodesics.

Finally, we have shown that there is an analogy between the spatial motion of photons in a dispersive medium and that of test particles subject to an external repulsion field. It is valid not only for electrostatic fields but also for any central potential. 

Let also note that even when we have applied the two types of optical metrics to the weak field regime, there is not any impediment to apply them to the strong gravitational region. For example, the 4-dimensional Gordon-like optical metric  \eqref{Gordon-metric-spherical} takes the form:
\begin{equation}\label{eq:gordonfr}
\begin{aligned}
\hat{g}_{\alpha\beta} dx^{\alpha} dx^{\beta}=&\frac{A(r)}{n^{2}(r)} dt^{2} - B(r) dr^{2} - C(r) d\Omega^2\\
=&\tilde{A}(r) dt^{2} - B(r) dr^{2} - C(r) d\Omega^2.
\end{aligned}
\end{equation}
As previously stated light rays which solve Eqs.\eqref{ham-jac} in the physical metric follows the same spatial orbits of null geodesics of this optical metric. As it is well known, circular null geodesics
which define photon spheres at radius $r=r_{\text{phot}}$ of a metric like \eqref{eq:gordonfr} satisfy the equation
\begin{equation}\label{ps}
\frac{C'(r)}{C(r)}\bigg|_{r=r_{\text{phot}}}=\frac{\tilde{A}'(r)}{\tilde{A}(r)}\bigg|_{r=r_{\text{phot}}},
\end{equation}
where a prime meaning a derivative with respect to $r$. Eq.\eqref{ps} follows from the study of null geodesics of \eqref{eq:gordonfr} or alternatively, in geometrical terms, from asking for the existence of a 3-dimensional timelike surface $S$ with the tracefree part of the second fundamental form vanishing (See \cite{Claudel:2000yi} for more details). 
The condition which follows from \eqref{ps} applied to the metric \eqref{eq:gordonfr}, namely
\begin{equation}\label{CnA}
C'(r)n(r)A(r)-A'(r)C(r)n(r)+2C(r)A(r)n'(r)=0
\end{equation}
agrees with the equation that follows from the Eq.(32) of Ref.\cite{Perlick:2015vta}, which was derived studying the timelike curves which are solutions of \eqref{ham-jac}.
It would be interesting to see how these results extend to the case of a rotating stationary spacetime. In such a situation, instead of an Riemannian optical metric that encodes the movement of particles, there is an associated Jacobi-Maupertuis-Randers-Finsler metric \cite{Gibbons-Jac-Maup-3}. In future works we will attack this more general situation.

\section*{Acknowledgments}

We would like to thank to Adam Rogers for his careful reading of the manuscript and suggestions. We acknowledge support from CONICET and SeCyT-UNC.

 \appendix

\section{Bending angle of a charged massive particle in a Reissner-Nordstr\"om spacetime from direct integration of the orbit equation}\label{app-RN}
In this Appendix we calculate the deflection of charged massive particles in the Reissner-Nordstr\"om spacetime directly from the spatial orbit equation. As we have seen in Section \ref{section-orbit-equation} the equation is given by,
 \begin{equation}\label{orbita-fisica-app}
    \bigg(\frac{d r}{d\varphi}\bigg)^2 = \frac{C(r)}{B(r)}\bigg( \frac{p_t^2}{p_\varphi^2} \frac{C(r)n^2(r)}{A(r)} -1 \bigg),
\end{equation}
where $A$, $B$ and $C$ are the metric components and $p_\varphi$ and $p_t$ are constants associated with the axial and temporal Killing vector fields, respectively.
In general one identifies these constant of motion with the angular momentum $J$ and energy $E$ of a test particle which measured by an asymptotic observer reduce to \eqref{Energy-angular-momentum}. Then,
 \begin{equation}\label{eq:ptph}
     \frac{p_t^2}{p_\varphi^2}=\frac{1}{v^2 b^2}.
 \end{equation}
 
On the other hand, for the Reissner-Nordstr\"om spacetime, we get
\begin{equation}
    A(r)=B^{-1}(r)=1-\frac{2m}{r}+\frac{Q^2}{r^2}, \ \ C(r)=r^2.
\end{equation}
Making the change of variable $u=1/r$ and taking into account the Eq.\eqref{eq:nparausar}, the orbit equation \eqref{orbita-fisica-app} reduces to,
\begin{equation}
\begin{aligned}
\bigg(\frac{du}{d\varphi}\bigg)^2=&\frac{1}{b^2}+\bigg(m(1-v^2)-\frac{qQ}{\mu}\sqrt{1-v^2}\bigg)\frac{2u}{b^2v^2}\\
&-\bigg(1-\frac{Q^2(1-v^2)}{b^2\mu^2v^2}(\mu^2-q^2)\bigg)u^2+2m u^3 \\&-Q^2 u^4,
\end{aligned}    
\end{equation}
If we want to calculate the deflection angle at the order $\mathcal{O}(\beta,\gamma,\delta,\beta^2,\gamma^2,\delta^2,\beta\delta)$ we have to solve the orbit equation at the same order. Then, let us consider the following ansatz in terms of the parameter defined in \eqref{parameters-dimless},
\begin{equation}
\begin{aligned}
u(\varphi)=&u_0 + u_1(\varphi)\beta+u_2(\varphi)\delta+u_3(\varphi)\gamma+u_4(\varphi)\beta\delta\\
&+u_5(\varphi)\beta^2+u_6(\varphi)\delta^2\\
&+\mathcal{O}(\gamma^2,\beta^3,\delta^3,\delta\beta^2,\beta\gamma,\beta\delta^2,\delta\gamma),
\end{aligned}    
\end{equation}
where $u_0=\frac{\sin\varphi}{b}$ and impose the asymptotic condition
\begin{equation}\label{eq:intcond}
    \lim_{\varphi\to 0}u_i(\varphi)=0.
\end{equation}
It follows that each of the  $u_i(\varphi)$ functions ($i=1..6$) must satisfy a differential equation of the following form:
\begin{equation}
 \cos\varphi \frac{d{u}_i(\varphi)}{d\varphi}+\sin\varphi u_i(\varphi)=F_i(\varphi,u_j(\varphi)),
\end{equation}
with $j\neq i$. Explicitly, the $F_i$ functions are given by:
\begin{widetext}
\begin{eqnarray}
F_1(\varphi)&=&\frac{\sin\varphi}{b}(\frac{1}{v^2}-\cos^2\varphi), \\
F_2(\varphi)&=&-\frac{\sin\varphi}{\mu v^2}\sqrt{1-v^2},\\
F_3(\varphi)&=&\frac{1}{2b}(\frac{\sin^2\varphi}{v^2}-\cos^4\varphi +3\cos^2\varphi-2),\\
F_4(\varphi,u_1(\varphi),u_2(\varphi))&=&u_2(\varphi)(3\sin^2\varphi+\frac{1}{v^2}-1)-b u_1(\varphi)\frac{\sqrt{1-v^2}}{\mu v^2}-b\bigg(u_1(\varphi)u_2(\varphi)+\frac{d{u}_1}{d\varphi}\frac{d{u}_2}{d\varphi}\bigg),\\
F_5(\varphi,u_1(\varphi))&=&u_1(\varphi)(2+\frac{1}{v^2}-3\cos^2\varphi)-\frac{b}{2}\bigg[u_1^2(\varphi)+\bigg(\frac{d{u}_1}{d\varphi}\bigg)^2\bigg],\\
F_6(\varphi,u_2(\varphi))&=&-\frac{b}{2\mu^2 v^2}\bigg\{\sin^2\varphi (1-v^2)+2\mu\sqrt{1-v^2}u_2(\varphi)+\bigg [u_2^2(\varphi)+\bigg(\frac{d{u}_2}{d\varphi}\bigg)^2\bigg]\mu^2 v^2\bigg\}.
\end{eqnarray}
\end{widetext}
Therefore, we can solve the first three equations for $u_1(\varphi),u_2(\varphi),u_3(\varphi)$, fixing the respective integration constants in order to satisfy the asymptotic condition \eqref{eq:intcond}, and after that we can solve the remaining equations for $u_4(\varphi),u_5(\varphi)$ and $u_6(\varphi)$ using the known expressions for $u_1(\varphi)$ and $u_2(\varphi)$ and fixing again the integration constants with the condition \eqref{eq:intcond}.

Doing so, the solution of the orbit equation in term of the physical parameters is given as follows,
\begin{widetext}
\begin{equation}
\begin{aligned}
u(\varphi)=&\frac{1}{b} \bigg[\sin\varphi + \bigg((1-\cos\varphi)+v^2(\cos^2\varphi-\cos\varphi)\bigg)\frac{m}{bv^2}+ \bigg( (2+v^2)\varphi\cos\varphi -2\sin\varphi-v^2\cos^2\varphi\sin\varphi \bigg)\frac{Q^2}{4b^2v^2} \\
&-\bigg(3v^2\cos^2\varphi\sin\varphi-[8\sin\varphi(1+v^2)-(12+3v^2)\varphi]\cos\varphi+2\sin\varphi(v^2-2) \bigg)\frac{m^2}{4b^2v^2}\\
&-(1-\cos\varphi)\sqrt{1-v^2}\frac{qQ}{bv^2\mu}-(\sin\varphi-3\varphi\cos\varphi+2\cos\varphi\sin\varphi)\sqrt{1-v^2}\frac{qQm}{b^2v^2\mu}\\
&-(\varphi\cos\varphi-\sin\varphi)(1-v^2)\frac{q^2Q^2}{2b^2v^2\mu^2}\bigg].
\end{aligned}
\end{equation}
\end{widetext}
At this point it important to remark that if we had calculated the deflection angle using the Gauss-Bonnet method we would only need to determine the function $u_1$ and $u_2$. In addition, we need  to compute the two-form $\mathcal{K}dS$ at the same order as required for the deflection angle, but this is easier than having to calculate the other functions $u_i$ ($i=3,4,..,6$). This is an operational advantage of the Gauss-Bonnet method that allows us to determine the deflection angle at the desired order in an easy way.

To finish, we may calculate the deflection angle $\alpha$ as follows,
\begin{equation}\label{angle-directly}
      \alpha \approx \tan\alpha=\lim_{\varphi\to\pi}\frac{\dot{y}(\varphi)}{\dot{x}(\varphi)},
\end{equation}
 where ``dot" means derivative respect to $\varphi$ and 
\begin{eqnarray}
    x(\varphi)&=&r(\varphi)\cos\varphi, \\
    y(\varphi)&=&r(\varphi)\sin\varphi.
\end{eqnarray}
Note that in principle one should take the limit of $\varphi$ going to $\pi+\alpha$ instead of $\pi$. However, it can be checked that if in Eq.\eqref{angle-directly} we take the limit $\varphi\to\pi+\Delta$, where $\Delta$ is assumed to be a quantity composed by terms of order $\beta,\gamma$, and $\delta$ they will contribute to the deflection angle as given by \eqref{angle-final-directly} with terms of higher order than considered, and therefore it is consistent to take $\varphi\to\pi$. 
Finally, using the relation \eqref{angle-directly} we get
\begin{widetext}
\begin{equation}\label{angle-final-directly}
\begin{aligned}
    \alpha=&\frac{2m}{b}\bigg(1+\frac{1}{v^2}\bigg)-\frac{2qQ}{bv^2}\frac{\sqrt{1-v^2}}{\mu}+\frac{3\pi m^2}{4b^2}\bigg(1+\frac{4}{v^2}\bigg)-\frac{\pi Q^2}{4b^2}\bigg( 1+\frac{2}{v^2}\bigg)-\frac{3\pi qQm}{b^2v^2}\frac{\sqrt{1-v^2}}{\mu}+\frac{\pi q^2 Q^2}{2b^2v^2}\frac{1-v^2}{\mu^2}. 
\end{aligned}    
\end{equation}
\end{widetext}
Using the first expression in \eqref{Energy-angular-momentum} for the energy of the particle we can see that the last expression for the deflection angle coincides with the previous one given in \eqref{alpha_RN-1}.



\begin{thebibliography}{10}
	
	\bibitem{Gallo-lens-2011}
	Emanuel Gallo and Osvaldo~M. Moreschi.
	\newblock Gravitational lens optical scalars in terms of energy-momentum
	distributions.
	\newblock {\em Phys. Rev. D}, 83:083007, 2011.
	
	\bibitem{Solar-Radio}
	D.~O. Muhleman and I.~D. Johnston.
	\newblock Radio propagation in the solar gravitational field.
	\newblock {\em Phys. Rev. Lett.}, 17:455--458, Aug 1966.
	
	\bibitem{Muhleman-1970}
	D.~O. Muhleman, R.~D. Ekers, and E.~B. Fomalont.
	\newblock Radio interferometric test of the general relativistic light bending
	near the sun.
	\newblock {\em Phys. Rev. Lett.}, 24:1377--1380, Jun 1970.
	
	\bibitem{Breuer-1980}
	J.~Breuer, R. A.;~Ehlers.
	\newblock Propagation of high-frequency electromagnetic waves through a
	magnetized plasma in curved space-time. i.
	\newblock {\em Proceedings Mathematical Physical \& Engineering Sciences}, 370,
	03 1980.
	
	\bibitem{Breuer-1981a}
	J.~Breuer, R. A.;~Ehlers.
	\newblock Propagation of high-frequency electromagnetic waves through a
	magnetized plasma in curved space-time. ii - application of the asymptotic
	approximation.
	\newblock {\em Proceedings of the Royal Society of London Series A}, 374, 65 (1981).
	
	\bibitem{Breuer-1981b}
	J.~Breuer, R. A.;~Ehlers.
	\newblock Propagation of electromagnetic waves through magnetized plasmas in
	arbitrary gravitational fields.
	\newblock {\em Astronomy \& Astrophysics}, 03 1981.
	
	\bibitem{Perlick-book}
	Volker Perlick.
	\newblock {\em Ray Optics, Fermat{'}s Principle, and Applications to General
		Relatively}.
	\newblock Lecture Notes in Physics 61. Springer-Verlag Berlin Heidelberg, first
	edition, 2000.
	
	\bibitem{BisnovatyiKogan:2008yg}
	G.~S. Bisnovatyi-Kogan and O.~{\relax Yu}. Tsupko.
	\newblock {Gravitational radiospectrometer}.
	\newblock {\em Grav. Cosmol.}, 15:20--27, 2009.
	
	\bibitem{BisnovatyiKogan:2010ar}
	G.~S. Bisnovatyi-Kogan and O.~{\relax Yu}. Tsupko.
	\newblock {Gravitational lensing in a non-uniform plasma}.
	\newblock {\em Mon. Not. Roy. Astron. Soc.}, 404:1790--1800, 2010.
	
	\bibitem{Tsupko:2013cqa}
	Oleg~Yu Tsupko and Gennady~S. Bisnovatyi-Kogan.
	\newblock {Gravitational lensing in plasma: Relativistic images at homogeneous
		plasma}.
	\newblock {\em Phys. Rev.}, D87(12):124009, 2013.
	
	\bibitem{Tsupko:2014sca}
	O.~{\relax Yu}. Tsupko and G.~S. Bisnovatyi-Kogan.
	\newblock {Influence of Plasma on Relativistic Images of Gravitational
		Lensing}.
	\newblock {\em Nonlin. Phenom. Complex Syst.}, 17(4):455--457, 2014.
	
	\bibitem{Tsupko:2014lta}
	O.~{\relax Yu}. Tsupko and G.~S. Bisnovatyi-Kogan.
	\newblock {Gravitational lensing in the presence of plasmas and strong
		gravitational fields}.
	\newblock {\em Grav. Cosmol.}, 20(3):220--225, 2014.
	
	\bibitem{Perlick:2015vta}
	Volker Perlick, Oleg~{\relax Yu}. Tsupko, and Gennady~S. Bisnovatyi-Kogan.
	\newblock {Influence of a plasma on the shadow of a spherically symmetric black
		hole}.
	\newblock {\em Phys. Rev.}, D92(10):104031, 2015.
	
	\bibitem{Bisnovatyi-Kogan:2015dxa}
	G.~S. Bisnovatyi-Kogan and O.~{\relax Yu}. Tsupko.
	\newblock {Gravitational Lensing in Plasmic Medium}.
	\newblock 2015.
	\newblock [Plasma Phys. Rep.41,562(2015)].
	
	\bibitem{Perlick:2017fio}
	Volker Perlick and Oleg~{\relax Yu}. Tsupko.
	\newblock {Light propagation in a plasma on Kerr spacetime: Separation of the
		Hamilton-Jacobi equation and calculation of the shadow}.
	\newblock {\em Phys. Rev.}, D95(10):104003, 2017.
	
	\bibitem{Bisnovatyi-Kogan:2017kii}
	Gennady Bisnovatyi-Kogan and Oleg Tsupko.
	\newblock {Gravitational Lensing in Presence of Plasma: Strong Lens Systems,
		Black Hole Lensing and Shadow}.
	\newblock {\em Universe}, 3(3):57, 2017.
	
	\bibitem{2013ApSS.346..513M}
	V.~S. {Morozova}, B.~J. {Ahmedov}, and A.~A. {Tursunov}.
	\newblock {Gravitational lensing by a rotating massive object in a plasma}.
	\newblock {\em Astrophysics and Space Science}, 346:513--520, 2013.
	
	\bibitem{2016ApSS.361..226A}
	A.~{Abdujabbarov}, B.~{Juraev}, B.~{Ahmedov}, and Z.~{Stuchl{\'{\i}}k}.
	\newblock {Shadow of rotating wormhole in plasma environment}.
	\newblock {\em Astrophysics and Space Science}, 361:226, 2016.
	
	\bibitem{2017IJMPD..2650051A}
	A.~{Abdujabbarov}, B.~{Toshmatov}, Z.~{Stuchl{\'{\i}}k}, and B.~{Ahmedov}.
	\newblock {Shadow of the rotating black hole with quintessential energy in the
		presence of plasma}.
	\newblock {\em International Journal of Modern Physics D}, 26:1750051--239,
	2017.
	
	\bibitem{2017IJMPD..2641011A}
	A.~{Abdujabbarov}, B.~{Toshmatov}, J.~{Schee}, Z.~{Stuchl{\'{\i}}k}, and
	B.~{Ahmedov}.
	\newblock {Gravitational lensing by regular black holes surrounded by plasma}.
	\newblock {\em International Journal of Modern Physics D}, 26:1741011--187,
	2017.
	
	\bibitem{2017PhRvD..96h4017A}
	A.~{Abdujabbarov}, B.~{Ahmedov}, N.~{Dadhich}, and F.~{Atamurotov}.
	\newblock {Optical properties of a braneworld black hole: Gravitational lensing
		and retrolensing}.
	\newblock {\em Phys. Rev. D.}, 96(8):084017, 2017.
	
	\bibitem{2018arXiv180203293T}
	B.~{Turimov}, B.~{Ahmedov}, A.~{Abdujabbarov}, and C.~{Bambi}.
	\newblock {Gravitational lensing by magnetized compact object in the presence
		of plasma}.
	\newblock {\em ArXiv: 1802.03293}, 2018.
	
	\bibitem{Rogers:2015dla}
	Adam Rogers.
	\newblock {Frequency-dependent effects of gravitational lensing within plasma}.
	\newblock {\em Mon. Not. Roy. Astron. Soc.}, 451(1):17--25, 2015.
	
	\bibitem{Rogers:2016xcc}
	Adam Rogers.
	\newblock {Escape and Trapping of Low-Frequency Gravitationally Lensed Rays by
		Compact Objects within Plasma}.
	\newblock {\em Mon. Not. Roy. Astron. Soc.}, 465(2):2151--2159, 2017.
	
	\bibitem{Rogers:2017ofq}
	Adam Rogers.
	\newblock {Gravitational Lensing of Rays through the Levitating Atmospheres of
		Compact Objects}.
	\newblock {\em Universe}, 3:3, 2017.
	
	\bibitem{2018MNRAS.475..867E}
	X.~{Er} and A.~{Rogers}.
	\newblock {Two families of astrophysical diverging lens models}.
	\newblock {\em Mon.Not.Roy.Astron.Soc.}, 475:867--878, March 2018.
	
	\bibitem{Crisnejo-gauss-bonnet-1}
	Gabriel Crisnejo and Emanuel Gallo.
	\newblock {Weak lensing in a plasma medium and gravitational deflection of
		massive particles using the Gauss-Bonnet theorem. A unified treatment}.
	\newblock {\em Phys. Rev.}, D97(12):124016, 2018.
	
	\bibitem{Er:2013efa}
	Xinzhong Er and Shude Mao.
	\newblock {Effects of plasma on gravitational lensing}.
	\newblock {\em Mon. Not. Roy. Astron. Soc.}, 437(3):2180--2186, 2014.
	
	\bibitem{Yan:2019etp}
	Haopeng Yan.
	\newblock {Influence of a plasma on the observational signature of a high-spin
		Kerr black hole}.
	\newblock {\em Phys. Rev. D} 99, 084050 (2019).
	
	\bibitem{Boero:2016nrd}
	Ezequiel~F Boero and Osvaldo~M Moreschi.
	\newblock Gravitational lens optical scalars in terms of energy-momentum
	distributions in the cosmological framework.
	\newblock {\em Mon. Not. Roy. Astron. Soc.}, 475(4):4683--4703, 2018.
	
	\bibitem{Crisnejo-lens-2018}
	Gabriel Crisnejo and Emanuel Gallo.
	\newblock Expressions for optical scalars and deflection angle at second order
	in terms of curvature scalars.
	\newblock {\em Phys. Rev. D}, 97:084010, 2018.
	
	\bibitem{Gallo-peculiar}
	Emanuel Gallo and Osvaldo Moreschi.
	\newblock {Peculiar anisotropic stationary spherically symmetric solution of
		Einstein equations}.
	\newblock {\em Mod. Phys. Lett.}, A27:1250044, 2012.
	
	\bibitem{Bozza:2015haa}
	V.~Bozza and A.~Postiglione.
	\newblock {Alternatives to Schwarzschild in the weak field limit of General
		Relativity}.
	\newblock {\em JCAP}, 1506(06):036, 2015.
	
	\bibitem{Gibbons-gauss-bonnet}
	G.~W. Gibbons and M.~C. Werner.
	\newblock {Applications of the Gauss-Bonnet theorem to gravitational lensing}.
	\newblock {\em Class. Quant. Grav.}, 25:235009, 2008.
	
	\bibitem{Jusufi:2015laa}
	Kimet Jusufi.
	\newblock {Gravitational lensing by Reissner-Nordstr{\"o}m black holes with
		topological defects}.
	\newblock {\em Astrophys. Space Sci.}, 361(1):24, 2016.
	
	\bibitem{Jusufi:2016wiz}
	Kimet Jusufi.
	\newblock {Light Deflection with Torsion Effects Caused by a Spinning Cosmic
		String}.
	\newblock {\em Eur. Phys. J.}, C76(6):332, 2016.
	
	\bibitem{Jusufi:2016sym}
	Kimet Jusufi.
	\newblock {Quantum effects on the deflection of light and the Gauss-Bonnet
		theorem}.
	\newblock {\em Int. J. Geom. Meth. Mod. Phys.}, 14(10):1750137, 2017.
	
	\bibitem{Jusufi:2017gyu}
	Kimet Jusufi.
	\newblock {Deflection angle of light by wormholes using the Gauss-Bonnet
		theorem}.
	\newblock {\em Int. J. Geom. Meth. Mod. Phys.}, 14(12):1750179, 2017.
	
	\bibitem{Ovgun:2018ran}
	Ali {\"O}vg{\"u}n, Kimet Jusufi, and Izzet Sakalli.
	\newblock {Gravitational lensing under the effect of Weyl and bumblebee
		gravities: Applications of Gauss{-}Bonnet theorem}.
	\newblock {\em Annals Phys.}, 399:193--203, 2018.
	
	\bibitem{Ovgun:2018fnk}
	Ali {\"O}vg{\"u}n.
	\newblock {Light deflection by Damour-Solodukhin wormholes and Gauss-Bonnet
		theorem}.
	\newblock {\em Phys. Rev.}, D98(4):044033, 2018.
	
	\bibitem{Jusufi:2018jof}
	Kimet Jusufi, Ali {\"O}vg{\"u}n, Joel Saavedra, Yerko V\'asquez, and P.~A.
	Gonz\'alez.
	\newblock {Deflection of light by rotating regular black holes using the
		Gauss-Bonnet theorem}.
	\newblock {\em Phys. Rev.}, D97(12):124024, 2018.
	
	\bibitem{Jusufi:2017uhh}
	Kimet Jusufi and Ali {\"O}vg{\"u}n.
	\newblock {Effect of the cosmological constant on the deflection angle by a
		rotating cosmic string}.
	\newblock {\em Phys. Rev.}, D97(6):064030, 2018.
	
	\bibitem{Jusufi:2017mav}
	Kimet Jusufi and Ali {\"O}vg{\"u}n.
	\newblock {Gravitational Lensing by Rotating Wormholes}.
	\newblock {\em Phys. Rev.}, D97(2):024042, 2018.
	
	\bibitem{Jusufi:2017vta}
	Kimet Jusufi, Ali {\"O}vg{\"u}n, and Ayan Banerjee.
	\newblock {Light deflection by charged wormholes in Einstein-Maxwell-dilaton
		theory}.
	\newblock {\em Phys. Rev.}, D96(8):084036, 2017.
	\newblock [Addendum: Phys. Rev.D96,no.8,089904(2017)].
	
	\bibitem{Jusufi:2017drg}
	Kimet Jusufi, Nayan Sarkar, Farook Rahaman, Ayan Banerjee, and Sudan Hansraj.
	\newblock {Deflection of light by black holes and massless wormholes in massive
		gravity}.
	\newblock {\em Eur. Phys. J.}, C78(4):349, 2018.
	
	\bibitem{Jusufi:2017xnr}
	Kimet Jusufi, Farook Rahaman, and Ayan Banerjee.
	\newblock {Semiclassical gravitational effects on the gravitational lensing in
		the spacetime of topological defects}.
	\newblock {\em Annals Phys.}, 389:219--233, 2018.
	
	\bibitem{Jusufi:2018waj}
	Kimet Jusufi.
	\newblock {Conical Morris-Thorne Wormholes with a Global Monopole Charge}.
	\newblock {\em Phys. Rev.}, D98(4):044016, 2018.
	
	\bibitem{Ishihara:2016vdc}
	Asahi Ishihara, Yusuke Suzuki, Toshiaki Ono, Takao Kitamura, and Hideki Asada.
	\newblock {Gravitational bending angle of light for finite distance and the
		Gauss-Bonnet theorem}.
	\newblock {\em Phys. Rev.}, D94(8):084015, 2016.
	
	\bibitem{Ishihara:2016sfv}
	Asahi Ishihara, Yusuke Suzuki, Toshiaki Ono, and Hideki Asada.
	\newblock {Finite-distance corrections to the gravitational bending angle of
		light in the strong deflection limit}.
	\newblock {\em Phys. Rev.}, D95(4):044017, 2017.
	
	\bibitem{Ovgun:2018xys}
	Ali {\"O}vg{\"u}n, Kimet Jusufi, and Izzet Sakalli.
	\newblock {Exact traversable wormhole solution in bumblebee gravity}.
	\newblock {\em Phys. Rev.}, D99(2):024042, 2019.
	
	\bibitem{Jusufi:2018kry}
	Kimet Jusufi.
	\newblock {Gravitational deflection of relativistic massive particles by Kerr
		black holes and Teo wormholes viewed as a topological effect}.
	\newblock {\em Phys. Rev.}, D98(6):064017, 2018.
	
	\bibitem{Jusufi:2017hed}
	Kimet Jusufi, Izzet Sakalli, and Ali {\"O}vg{\"u}n.
	\newblock {Effect of Lorentz Symmetry Breaking on the Deflection of Light in a
		Cosmic String Spacetime}.
	\newblock {\em Phys. Rev.}, D96(2):024040, 2017.
	
	\bibitem{Jusufi:2017lsl}
	Kimet Jusufi, Marcus~C. Werner, Ayan Banerjee, and Ali {\"O}vg{\"u}n.
	\newblock {Light Deflection by a Rotating Global Monopole Spacetime}.
	\newblock {\em Phys. Rev.}, D95(10):104012, 2017.
	
	\bibitem{Ovgun:2018tua}
	Ali {\"O}vg{\"u}n, Izzet Sakalli, and Joel Saavedra.
	\newblock {Shadow cast and Deflection angle of Kerr-Newman-Kasuya spacetime}.
	\newblock {\em JCAP}, 1810(10):041, 2018.
	
	\bibitem{Javed:2019qyg}
	Wajiha Javed, Rimsha Babar, and Ali {\"O}vg{\"u}n.
	\newblock {The effect of the Brane-Dicke coupling parameter on weak
		gravitational lensing by wormholes and naked singularities}.
	\newblock {\em Phys. Rev.}, D99(8):084012, 2019.
	
	\bibitem{deLeon:2019qnp}
	Karlo de~Leon and Ian Vega.
	\newblock {Weak gravitational lensing by two-power-law densities using the
		Gauss-Bonnet theorem}.
	\newblock {\em Phys.
Rev. D} 99, 124007 (2019).
	
	\bibitem{Haroon:2018ryd}
	Sumarna Haroon, Mubasher Jamil, Kimet Jusufi, Kai Lin, and Robert~B. Mann.
	\newblock {Shadow and Deflection Angle of Rotating Black Holes in Perfect Fluid
		Dark Matter with a Cosmological Constant}.
	\newblock {\em Phys. Rev.}, D99(4):044015, 2019.
	
	\bibitem{Jusufi:2018gnz}
	Kimet Jusufi, Ayan Banerjee, Galin Gyulchev, and Muhammed Amir.
	\newblock {Distinguishing rotating naked singularities from Kerr-like wormholes
		by their deflection angles of massive particles}.
	\newblock {\em Eur. Phys. J.}, C79(1):28, 2019.
	
	\bibitem{Crisnejo:2018ppm}
	Gabriel Crisnejo, Emanuel Gallo, and Adam Rogers.
	\newblock {Finite distance corrections to the light deflection in a
		gravitational field with a plasma medium}.
	\newblock {\em Phys. Rev. D} 99, 124001, (2019).
	
	\bibitem{Weyl-1917}
	Hermann Weyl.
	\newblock Zur gravitationstheorie.
	\newblock {\em Annalen der Physik}, 359, 1917.
	
	\bibitem{Schulze-Koops:2017tkc}
	Karen Schulze-Koops, Volker Perlick, and Dominik~J. Schwarz.
	\newblock {Sachs equations for light bundles in a cold plasma}.
	\newblock {\em Class. Quant. Grav.}, 34(21):215006, 2017.
	
	\bibitem{W-1923}
	W.~Gordon.
	\newblock Zur lichtfortpflanzung nach der relativit\"atstheorie.
	\newblock {\em Annalen der Physik}, 377, 1923.
	
	\bibitem{Kulsrud-1992}
	Russell Kulsrud and Abraham Loeb.
	\newblock Dynamics and gravitational interaction of waves in nonuniform media.
	\newblock {\em Phys. Rev. D}, 45:525--531, Jan 1992.
	
	\bibitem{Gibbons-Jacobi-static}
	G.~W. Gibbons.
	\newblock {The Jacobi-metric for timelike geodesics in static spacetimes}.
	\newblock {\em Class. Quant. Grav.}, 33(2):025004, 2016.
	
	\bibitem{Gibbons-Jac-Maup-1}
	Sumanto Chanda, G.~W. Gibbons, and Partha Guha.
	\newblock {Jacobi-Maupertuis-Eisenhart metric and geodesic flows}.
	\newblock {\em J. Math. Phys.}, 58(3):032503, 2017.
	
	\bibitem{Gibbons-Jac-Maup-2}
	Sumanto Chanda, G.~W. Gibbons, and Partha Guha.
	\newblock {Jacobi{-}Maupertuis metric and Kepler equation}.
	\newblock {\em Int. J. Geom. Meth. Mod. Phys.}, 14(07):1730002, 2017.
	
	\bibitem{Gibbons-Jac-Maup-3}
	Sumanto Chanda, G.~W. Gibbons, Partha Guha, Paolo Maraner, and Marcus~C.
	Werner.
	\newblock {Jacobi-Maupertuis Randers-Finsler metric for curved spaces and the
		gravitational magnetoelectric effect}.
	\newblock Arxiv: 1903.11805 (2019).
	
	\bibitem{Das:2016opi}
	Praloy Das, Ripon Sk, and Subir Ghosh.
	\newblock {Motion of charged particle in Reissner{-}Nordstr{\"o}m spacetime: a
		Jacobi-metric approach}.
	\newblock {\em Eur. Phys. J.}, C77(11):735, 2017.
	
	\bibitem{book:75670}
	John~Lighton Synge.
	\newblock {\em Relativity: The general theory}.
	\newblock North Holland Publishing Co., 1960.
	
	\bibitem{Bartelmann:2010fz}
	Matthias Bartelmann.
	\newblock {Gravitational Lensing}.
	\newblock {\em Class. Quant. Grav.}, 27:233001, 2010.
	
	\bibitem{Seitz:1994xf}
	Stella Seitz, Peter Schneider, and Jurgen Ehlers.
	\newblock {Light propagation in arbitrary space-times and the gravitational
		lens approximation}.
	\newblock {\em Class. Quant. Grav.}, 11:2345--2374, 1994.
	
	\bibitem{Frittelli:2000bc}
	Simonetta Frittelli, Thomas~P. Kling, and Ezra~T. Newman.
	\newblock {Image distortion from optical scalars in nonperturbative
		gravitational lensing}.
	\newblock {\em Phys. Rev.}, D63:023007, 2000.
	
	\bibitem{Pang:2018jpm}
	Xiankai Pang and Junji Jia.
	\newblock {Gravitational lensing of massive particles in Reissner-Nordstr{\"o}m
		spacetime}.
	\newblock {\em Class. Quant. Grav.}, 36(6):065012, 2019.
	
	\bibitem{book:synge}
	John~Lighton Synge.
	\newblock {\em Relativity: The special theory}.
	\newblock North Holland Publishing Co., 1956.
	
	\bibitem{Claudel:2000yi}
	Clarissa-Marie Claudel, K.~S. Virbhadra, and G.~F.~R. Ellis.
	\newblock {The Geometry of photon surfaces}.
	\newblock {\em J. Math. Phys.}, 42:818--838, 2001.
	
\end{thebibliography}


\end{document}